\documentclass[universe,article,accept,moreauthors]{Definitions/mdpi} 
\usepackage{amsmath,amssymb,bm}

\usepackage{graphicx}
\usepackage{booktabs}
\usepackage{array}
\usepackage{xcolor}
\usepackage{caption}
\usepackage{float}

\firstpage{1} 
\pubvolume{12}
\issuenum{10}
\articlenumber{309}
\pubyear{2026}
\copyrightyear{2026}
\externaleditor{Anzhong Wang} %
\datereceived{20 August 2026} 
\daterevised{12 October 2026} %
\dateaccepted{14 October 2026} 
\datepublished{16 October 2026} 
\hreflink{https://doi.org/10.3390/universe12100309} %

\Title{Spin-Inversion Degeneracies in Restricted Inspiral Waveforms for LISA}

\Author{Kata Karácsonyi %
 $^{1,2}$ and László Árpád Gergely $^{1,2,}$*}

\AuthorNames{Kata Karácsonyi and László Árpád Gergely}

\address{%
$^{1}$ \quad Department of Theoretical Physics, HUN-REN Wigner Research Centre for Physics, Konkoly-Thege Miklós út~29--33, 1121 Budapest, Hungary; karacsonyi.kata@wigner.hu %
\\
$^{2}$ \quad Department of Theoretical Physics, University of Szeged,  Tisza Lajos kör\'ut 84--86, 6720 Szeged, Hungary\\}

\corres{Correspondence: gergely.laszlo@wigner.hu}

\abstract{Spin inversions appear in several settings. Analytical work predicts a single spin flip during unequal-mass supermassive binary inspiral, while numerical relativity and post-Newtonian calculations show repeated flip-flop motion in comparable-mass binaries. Secular spin evolution also predicts additional cases driven by spin-induced mass quadrupoles. Whether these effects can be distinguished in gravitational-wave data is still unclear. We combine the secular spin angle equations with a quasi-circular second post-Newtonian frequency evolution and build a restricted waveform weighted by the sky-averaged LISA sensitivity. We study five near-equal-mass injections with detector-frame total mass $M=2\times10^5M_\odot$, including Kerr flip-flops and one quadrupole-induced case. Each injection is compared with physically evolving waveforms constrained to have no orbital-plane crossings. We search all four no-inversion sectors, vary the masses, spin magnitudes, and initial spin angles, and maximize over time, phase, and overall amplitude. Large spin motion does not by itself lead to a clearly different waveform in this restricted model. We compare configurations with different spin magnitudes, relative spin orientations, and angular excursions at eleven observer inclinations between $0^\circ$ and $90^\circ$. With each injection normalized to signal SNR $\rho_\star=100$, four of the five cases leave a residual SNR below unity for the $0^\circ$ viewing direction along the total angular momentum. At nonzero inclinations, the residual associated with the largest match found reaches $7.07$ in the two-active-spin case at $55^\circ$. These upper bounds on the minimum residual within the allowed template family suggest potential distinguishability of spin inversions and motivate studies with more complete waveforms and the full LISA response.}


\begin{document}
\section{Introduction}
Two distinct spin-inversion mechanisms motivate this study. First, analytical investigations of supermassive black-hole (SMBH) binaries established that the primary spin can undergo a large, non-recurrent reorientation during the early inspiral, accessible to post-Newtonian (PN) theory~\cite{GergelyBiermann2009,GergelyBiermannCaramete2010}. Such long-lived, slowly evolving supermassive systems are primary targets for the Laser Interferometer Space Antenna (LISA), a planned space-based gravitational-wave observatory operating in the millihertz band~\cite{LISA2023}.  Second, numerical relativity (NR) simulations of strongly precessing, equal-mass black-hole (BH) binaries uncovered a periodic “flip-flop” motion; while NR tracked this evolution through roughly half a cycle, its accompanying 3.5PN continuation predicted that the spin axis would flip-flop repeatedly throughout the remaining inspiral~\cite{LoustoHealy2015}. Subsequent equal-mass NR calculations confirmed this generic strong-field behavior, while their unequal-mass extensions continued to rely on 3.5PN %
evolution~\cite{LoustoHealyNakano2016}. A later analytical study of the closed secular spin system successfully reproduced these large flip-flops of the secondary spin when the dominant spin is nearly coplanar with the orbit, while uncovering a third dynamical possibility: a new class of quadrupole-induced flip-flops driven by specific compact-object mass quadrupoles, such as a dominant-spin neutron star with a quadrupole coefficient \(w_1\simeq3\)~\cite{KeresztesTapaiGergely2021}. Consequently, non-recurrent spin flips, recurrent strong-field flip-flops, and quadrupole-induced inversions are well-established theoretical predictions. Whether any of these geometric modulations can be successfully identified in gravitational-wave observations remains unknown. These physical mechanisms operate across fundamentally distinct mass-ratio regimes. The analytical SMBH spin flip occurs at intermediate mass ratios, \(1/30 \lesssim q = m_2/m_1 \lesssim 1/3\), or inverse ratio between 3 and 30. An earlier phenomenological estimate, based on an observationally motivated broken-power-law SMBH mass function and a mass-dependent galaxy-encounter weighting, found that approximately \(48.9\%\) of the weighted encounters fall within this mass-ratio range~\cite{GergelyBiermann2012}. Modern cosmological simulations continue to predict substantial populations of unequal-mass massive-black-hole mergers in this interval, while showing that their relative abundance depends on the black-hole mass range, redshift, seeding prescription, and merger-delay model~\cite{WangEtAl2025,LiaoEtAl2025}. While nearly equal-mass mergers require two comparatively rare massive BHs, extreme mass-ratio binaries sink inefficiently via dynamical friction and may fail to coalesce within a Hubble time~\cite{SesanaVolonteriHaardt2007}. Although recent cosmological calculations predict a broader distribution, this intermediate interval remains a strongly motivated astrophysical prior rather than an absolute boundary~\cite{SayebEtAl2021}. By contrast, NR flip-flop simulations have focused strictly on equal masses (\(q=1\)). Recurrent flip-flop evolution and its unequal-mass extensions rely on 3.5PN integrations where the associated unstable sector is tightly concentrated at \(0.5 < q < 1\)~\cite{LoustoHealy2015,LoustoHealyNakano2016,LoustoHealy2016}. This study deliberately tests this near-equal-mass sector.

The absolute total mass and the mass ratio play different roles in the spin-inversion mechanisms. After expressing the PN evolution in terms of $\tau=t/M$ and $v=(\pi Mf)^{1/3}$, rescaling $M$ at fixed $q$, spin magnitudes, quadrupole parameters, initial angles, and the same $v$ interval changes the physical time and frequency scales but not the dimensionless spin trajectory. By contrast, the non-recurrent dominant-spin flip is intrinsically mass-ratio-dependent. With $q=m_2/m_1\leq1$, the leading-order ratio $L_N/S_1=q/(\chi_1 v)$, so radiation reaction can drive the system through $L_N\approx S_1$ only for suitable $q$, spin magnitude, and initial geometry. Recurrent Kerr flip-flops instead arise from the coupled two-spin dynamics for appropriate configurations, while quadrupole-induced flip-flops occur when the spin-induced quadrupole terms alter the secular dynamics. Thus, the label SMBH refers to the astrophysical setting of the single spin-flip mechanism and its LISA application. Generally, there is no intrinsic requirement on the absolute total mass. Instead, the significant quantity is the ratio of component masses.

The instantaneous dynamical state of the binary is defined by the relative configuration of the angular-momentum triad %
 \((\bm{S}_1, \bm{S}_2, \bm{L}_N)\). At fixed magnitudes, the scalar products
$\cos\kappa_1$, $\cos\kappa_2$, and $\cos\gamma$
determine the three pairwise angles. To specify the signed relative
azimuth, one must also retain the sign of
$\sin\Delta\zeta$. Because nomenclature varies across the literature, we use spin inversion as a broad term for any evolution in which a
signed spin projection changes sign. An inversion occurs
when $\cos\kappa_i$ crosses zero. The size of the angular
motion is characterized by
$\Delta\kappa_i=\kappa_{i,\max}-\kappa_{i,\min}$. Within the 2PN model used here, the spin geometry is governed by spin--orbit, spin--spin, and quadrupole--monopole couplings, together with gravitational radiation dissipation.

\textls[-15]{The observational footprint of these geometric spin couplings is indirect, as gravitational-wave detectors cannot track the physical orientation of the angular momentum vectors directly, but instead record their signatures through the resulting modulations in the data stream. They modify the predicted detector strain through accumulated phase, precession-induced modulations, and time-dependent polarization structures. Because a gravitational-wave instrument records a raw strain time series rather than the underlying spin vectors or orbital trajectories, source dynamics should be inferred through comparison with waveform templates. A weak spin may cross the orbital plane while contributing negligibly to the overall signal, whereas the reorientation of a dynamically dominant spin can produce a substantial accumulated phase change. Observational distinguishability therefore depends not simply on the
angular motion, but on whether the resulting restricted waveform can be
reproduced by another physically evolving no-inversion binary with
different source parameters. Whether this degeneracy persists for
complete precessing waveforms remains to be determined. A recent LISA-oriented study illustrated spin-flip waveform effects by manually varying spin polar angles in an approximant without explicit precession~\cite{PislanEtAl2026}. The secular spin dynamics of eccentric binaries with spin--orbit, spin--spin, and quadrupole--monopole interactions was previously derived as a closed autonomous system through 2PN order~\cite{KeresztesTapaiGergely2021}. Analytical limits of this system reproduce the known large flip-flop behavior and reveal the new quadrupole-induced case, whose fixed points and stability depend strongly on the compact-object quadrupoles~\cite{KeresztesGergely2021}. Independently, the frequency evolution and accumulated cycles of spinning binaries with mass quadrupoles and magnetic dipoles were assembled through 2PN order in Ref.~\cite{MikocziVasuthGergely2005}, including individual-spin self-interaction terms further discussed in Ref.~\cite{GergelyMikoczi2009}.}

The calculations in this work treat the binary as an isolated vacuum system. Real massive-black-hole binaries may interact with gaseous discs, stellar backgrounds, or dark-matter distributions. Such environments can modify the orbital evolution and hence the waveform, while gas accretion and disc torques can also alter black-hole spin orientations~\cite{BarausseCardosoPani2014,SteinleGerosa2023,GargEtAl2024}. These effects are not included in the present 2PN vacuum dynamics. Incorporating them into the orbital and spin evolution and into the waveform is left to future studies. Here, we isolate the spin-inversion dynamics and its degeneracy within the restricted vacuum model. In this work, we combine the circular limit of the secular angle equations with the 2PN frequency evolution equations, setting magnetic dipoles to zero while allowing the spin-induced quadrupole coefficients \(w_{i}\) to take either their Kerr values or chosen non-Kerr values. We first evaluate Kerr spin-inversion trajectories and then the predicted quadrupole-induced case near \(w_1 \simeq 3\). In each case, we search all possible no-inversion sectors while varying the
total mass, mass ratio, spin magnitudes, and initial spin angles, and while
maximizing over time, phase, and overall amplitude. We include the $0^\circ$ viewing direction along the total angular momentum and repeat the no-inversion comparison at eleven observer inclinations, reconstructing the orbital azimuth needed for the restricted amplitude modulation. We use geometrized units $G=c=1$.

\section{Compact Binary Dynamics and Waveform Modeling}
\subsection{Spin Dynamics}

All masses %
entering the waveform are detector-frame redshifted masses. We denote the mass components such that $m_1\geq m_2$, thus defining the total mass, the mass ratio and the symmetric mass ratio as %
\begin{equation}
 M=m_1+m_2\ ,\qquad q=\frac{m_2}{m_1}\leq1\ ,\qquad
 \eta=\frac{m_1m_2}{M^2}=\frac{q}{(1+q)^2}\ .
\end{equation}
The %
component masses $m_i$ are held fixed during each evolution. Radiation reaction changes the orbital energy and angular momentum; accretion-driven mass evolution and environmental mass exchange are not included.
The frequency $f$ is that of the dominant $m=2$ gravitational-wave harmonic, so that the orbital angular frequency is $\Omega=\pi f$. The PN velocity variable and dimensionless time are
\begin{equation}
 v=(M\Omega)^{1/3}=(\pi Mf)^{1/3}\ ,\qquad \tau=\frac{t}{M}\ .
 \label{Eq v and tau}
\end{equation}

Ordinary precession changes where the compact binary's intrinsic angular momentum points in space. A spin inversion changes which side of the orbital plane it points toward. The spin's magnitude is written as
\begin{equation}
 S_i=\chi_i m_i^2\ ,
\end{equation}
where $i=1,2$ and $\chi_i\in[0,1]$ are the dimensionless spin magnitudes for a Kerr BH. Let $\bm L_N$ denote the Newtonian orbital angular momentum, normal to the instantaneous orbital plane. The total angular momentum is approximately
\begin{equation}
    \bm J \simeq \bm L_N +\bm S_1 + \bm S_2\ .
    \label{Eq total angular momentum}
\end{equation}
During the early inspiral, gravitational radiation decreases $\bm J$ slowly. On the shorter precession timescale, we may first treat $\bm J$ as nearly constant. Another idea is to neglect the radiation reaction, so that $d\bm J/dt=0$, in which case it is evident from Equation \eqref{Eq total angular momentum} that, when the spins turn, $\bm L_N$ must also turn. This produces two distinct types of motion: azimuthal precession (the spins and $\bm L_N$ rotate around the nearly fixed $\bm J$) and nutation or polar motion (the angles between the spins and $\bm L_N$ change). Only the latter motion may produce a spin~inversion.

We define
\begin{equation}
    \cos\kappa_i=\hat{\bm S}_i\cdot\hat{\bm L}_N\ ,
\end{equation}
where $\kappa_i$ helps to describe the orientation of $\hat{\bm S}_i$ ($\kappa_i=0$: aligned with $\bm L_N$; $\kappa_i=\pi/2$: in the orbital plane; $\kappa_i=\pi$: anti-aligned with $\bm L_N$) and we may define the relative spin angle
\begin{equation}
    \cos \gamma=\hat{\bm S}_1\cdot\hat{\bm S}_2\ .
\end{equation}
If $\zeta_i$ are the azimuthal angles of the spins around $\bm L_N$, then we may define the azimuthal difference
\begin{equation}
    \Delta \zeta=\zeta_1-\zeta_2 \ ,
\end{equation}
and the expression
\begin{equation}
    \cos\gamma=\cos\kappa_1\cos\kappa_2+\sin\kappa_1\sin\kappa_2\cos\Delta\zeta \ ,
    \label{Eq gamma}
\end{equation}
therefore, the internal spin configuration can be described by the three angles: $\kappa_1, \kappa_2$ and $\Delta\zeta$. These angles are independent of the binary's relative orientation to the observer.

In PN theory, the spin evolution has the form~\cite{Kidder1995}
\begin{equation}
    \frac{d\bm S_i}{dt}=\bm\Omega_i\times\bm S_i\ ,
\end{equation}
where $\bm\Omega_i$ is the precession angular velocity, which depends on $\bm L_N$, the other spin, the component masses, and the orbital separation. The cross-product form expresses conservative precession with fixed spin magnitude and is not specific to a PN expansion. In the present calculation, however, the precession angular velocity $\bm\Omega_i$ is the PN-expanded one. It is straightforward to show that the cross product gives $S_i=\text{constant}$ in conservative spin~precession.

The principal interactions are spin--orbit coupling (from 1.5PN order), spin--spin coupling (from 2PN order), and quadrupole--monopole coupling (from 2PN order). For equal or nearly equal masses, the spin--orbit parts of the precession rates of two nonzero spins are similar. The spin--spin interaction can then be especially important for the relative polar motion. For unequal masses, differences between the spin--orbit precession rates also contribute. Ordinary precession does not necessarily result in a spin inversion. Spin inversion is only confirmed when the spin inverts relative to the orbit, meaning that $\bm S_i\cdot\bm L_N=S_iL_N\cos\kappa_i$ changes sign. Equivalently, $\kappa_i$ crosses $\pi/2$~\cite{ApostolatosEtAl1994}. Therefore, to decide whether a spin inversion has occurred, one should examine $\cos\kappa_i$.

Suppose $S_1>S_2$; the smaller spin may change direction by a large angle, while the larger spin may change by a smaller angle, leaving $\bm S$ stable. Here, $\bm S=\bm S_1+\bm S_2$ denotes the total spin. Let $\lambda_2$ be the angle between $\bm S_2$ and $\bm S$; then,
\begin{equation}
    \cos \lambda_2 = \frac{S^2+S_2^2-S_1^2}{2SS_2}\ .
\end{equation}
If $S_1>S_2$, it is possible to have a large $\lambda_2$ but a small $\lambda_1$. Thus, the smaller spin can sweep through a much larger angle than the larger spin. Consequently, complete flip-flops most naturally occur in the smaller of two unequal spin vectors. A spin flip-flop is a repeated oscillation between the limiting orientations $\kappa_{i,\text{min}}\leftrightarrow \kappa_{i,\text{max}}$. Its angular difference can be defined as
\begin{equation}
    \Delta \kappa_i =\kappa_{i,\text{max}}-\kappa_{i,\text{min}}\ ,
\end{equation}
therefore, for a complete flip-flop, $\Delta \kappa_i\approx\pi$. Thus, the same BH's spin can evolve as \textsc{aligned}$\leftrightarrow$\textsc{anti-aligned}~\cite{LoustoHealyNakano2016}.

\subsection{Evolution of the Internal Angles \texorpdfstring{$\kappa_1, \kappa_2,\Delta \zeta$}{kappa1, kappa2, Delta zeta}}
\label{subsec:internal-angle-evolution}

We use the parameter
\begin{equation}
 x_i=\chi_i v\ ,
 \label{Eq x parameter}
\end{equation}
in which $v$, given by Equation \eqref{Eq v and tau}, measures how relativistic and compact the orbit is. As the binary inspirals, $f$ and $v$ increase, so $x_i$ also increases even though $\chi_i$ remains constant. We parametrize the spin-induced mass quadrupole of the component $i$ by
\begin{equation}
 Q_i=-w_i\frac{S_i^2}{m_i}\ .
 \label{Eq spin-quadrupole}
\end{equation}
For each trajectory, $w_i$ is held constant. Varying $w_i$ between models represents different spin-induced quadrupole structures of the compact object (with $w_i=1$ for a Kerr black hole), not a time evolution of the coefficient for a single body.
The two polar angles obey
\begin{align}
 \frac{d\kappa_1}{d\tau}={}&\frac{3}{2}\eta v^5
 \left(1+q-x_1\cos\kappa_1-qw_2x_2\cos\kappa_2\right)
 x_2\sin\kappa_2\sin\Delta\zeta\ ,
 \label{Eq kappa derivatives1}\\
 \frac{d\kappa_2}{d\tau}={}&-\frac{3}{2}\eta v^5
 \left(1+q^{-1}-x_2\cos\kappa_2-q^{-1}w_1x_1\cos\kappa_1\right)
 x_1\sin\kappa_1\sin\Delta\zeta\ .
 \label{Eq kappa derivatives2}
\end{align}
Equations \eqref{Eq kappa derivatives1} and \eqref{Eq kappa derivatives2} show which spin drives each nutation: $d\kappa_1/d\tau$ is proportional to $x_2$, whereas $d\kappa_2/d\tau$ is proportional to $x_1$; therefore, each spin may drive the polar motion of the other. Thus, a dominant spin can produce a large reorientation of the smaller spin while changing its own polar angle much less. If $\chi_1\gg\chi_2$, then $x_1\gg x_2$; therefore, we expect $|d\kappa_2/d\tau|\gg|d\kappa_1/d\tau|$. Equations \eqref{Eq kappa derivatives1} and \eqref{Eq kappa derivatives2} both contain $\sin\Delta\zeta$; therefore, when $\Delta\zeta=0$ or $\pi$, the projections of the two spins onto the orbital plane are parallel or anti-parallel. The three vectors are then coplanar.

The relative azimuth angle obeys
\begin{equation}
 \frac{d\Delta\zeta}{d\tau}=\frac{3}{2}\eta v^5\,\mathcal F\ ,
 \label{Eq deltazeta derivative}
\end{equation}
with
\begin{equation}
 c_i=\cos\kappa_i\ ,\qquad
 s_i=\sin\kappa_i\ ,\qquad
 C=\cos\Delta\zeta\ .
\end{equation}
The complete circular-limit expression is
\begingroup\makeatletter\def\f@size{9}\check@mathfonts
\def\maketag@@@#1{\hbox{\m@th\normalsize \normalfont#1}}%
\begin{align}
 \mathcal F={}&q-q^{-1}
 +\left(1+2q^{-1}-w_1-q^{-1}w_1x_1c_1\right)x_1c_1
-\left(1+2q-w_2-qw_2x_2c_2\right)x_2c_2
 \nonumber\\
 &-\left(1+q^{-1}-q^{-1}w_1x_1c_1\right)
 x_1\frac{c_2}{s_2}s_1C+\left(1+q-qw_2x_2c_2\right)
 x_2\frac{c_1}{s_1}s_2C
-x_1x_2
 \left(\frac{s_2}{s_1}-\frac{s_1}{s_2}\right)C\ .
 \label{Eq deltazeta's F}
\end{align}
\endgroup
One can verify that, for equal mass configurations, the leading difference between the two spin--orbit precession rates disappears, due to the terms $q-q^{-1}$. Equations~\eqref{Eq kappa derivatives1}--\eqref{Eq deltazeta derivative} give the derivatives of the three internal angles for prescribed values of $v$, $m_i$, $\chi_i$, and $w_i$. They form the circular limit of the orbit-averaged 2PN spin-angle system derived by Keresztes, T\'apai \& Gergely~\cite{KeresztesTapaiGergely2021}, obtained by orbit-averaging the spin-precession dynamics over the fast orbital motion and projecting the resulting vector evolution onto the polar angles $\kappa_i$ relative to $\bm L_N$ and the relative azimuth $\Delta\zeta$. The leading terms in the polar equations scale as $\eta\chi_jv^6$. The terms proportional to $(1+q)x_2$ and $(1+q^{-1})x_1$ are the leading spin--orbit contributions. The terms proportional to $x_1$ and $x_2$ arise from the mutual spin--spin interaction, while those proportional to $w_ix_i^2$ arise from the quadrupole--monopole interaction.

At $\kappa_i=0$ or $\pi$, the projection of $\bm S_i$ onto the orbital plane vanishes and its azimuth is undefined. The spin dynamics itself remains regular. Near a generic passage through either polar axis, the local solution has $\cos\Delta\zeta=\mathcal{O}(\delta\kappa_i)$ and $\sin\kappa_i=\mathcal{O}(|\delta\kappa_i|)$, where $\delta\kappa_i$ is the angular distance from the pole. Hence, the combinations $\cos\Delta\zeta/\sin\kappa_i$ remain finite. The azimuth changes by $\pi$ when the spin passes through the axis. The spin-inversion condition $\kappa_i=\pi/2$ is regular. The large-flip-flop limit follows directly when $\chi_1$ dominates; in such a~case,
\begin{equation}
 \frac{\mathcal{O}(d\kappa_1/d\tau)}{\mathcal{O}(d\kappa_2/d\tau)}
 \sim q\frac{\chi_2}{\chi_1}\ll1\ .
 \label{eq:kappa-rate-ratio}
\end{equation}
At leading order in $\chi_2/\chi_1$, the remaining equations are~\cite{KeresztesTapaiGergely2021}
\begin{align}
 \frac{d\kappa_2}{d\tau}&\simeq\frac{3}{2}\eta v^5
 \mathcal B\sin\Delta\zeta\ ,
 \label{Eq weakspin kappa}\\
 \frac{d\Delta\zeta}{d\tau}&\simeq\frac{3}{2}\eta v^5
 \left(\mathcal A+\mathcal B\cot\kappa_2\cos\Delta\zeta\right)\ ,
 \label{Eq weakspin zeta}
\end{align}
where
\begin{align}
 \mathcal A={}&q-q^{-1}
 +\left(1+2q^{-1}-w_1-q^{-1}w_1x_1\cos\kappa_1\right)
 x_1\cos\kappa_1\ ,
 \label{Eq weakspin A}\\
 \mathcal B={}&-\left(1+q^{-1}-q^{-1}w_1x_1\cos\kappa_1\right)
 x_1\sin\kappa_1\ .
 \label{Eq weakspin B}
\end{align}
To display the oscillation, we hold $v$ and $\kappa_1$ constant over one precession cycle and introduce
\begin{equation}
 d\lambda=\frac{3}{2}\eta v^5d\tau\ ,
 \qquad Y=\sin\kappa_2\sin\Delta\zeta\ .
\end{equation}
Through Equations~\eqref{Eq weakspin kappa} and \eqref{Eq weakspin zeta}, one finds the harmonic oscillator equation describing \mbox{the flip-flop:}
\begin{equation}
 \frac{d^2Y}{d\lambda^2}+\Omega_{\rm ff}^2Y=0\ ,
 \qquad \Omega_{\rm ff}=\sqrt{\mathcal A^2+\mathcal B^2}\ .
 \label{Eq flipflop-oscillator}
\end{equation}
The solution is
\begin{equation}
 Y=K_1\cos(\Omega_{\rm ff}\lambda+D)\ ,
\end{equation}
where $K_1$ and $D$ are fixed by the initial angles. The parameter $\Omega_{\rm ff}$ is the flip-flop frequency with respect to the rescaled time $\lambda$. With respect to $\tau$, the corresponding frequency is
\begin{equation}
    \omega_{\rm ff}^{\tau}=\frac{3}{2}\eta v^5\Omega_{\rm ff}\ .
\end{equation}
The polar equation with respect to $\lambda$ can be written as
\begin{equation}
 \frac{d\cos\kappa_2}{d\lambda}=-\mathcal B Y\ .
\end{equation}
Integrating the harmonic solution gives
\begin{equation}
 \cos\kappa_2=K_2-\frac{\mathcal B K_1}{\Omega_{\rm ff}}
 \sin(\Omega_{\rm ff}\lambda+D)\ .
 \label{Eq weakspin-solution}
\end{equation}
Hence, the peak-to-peak range is
\begin{equation}
 (\cos\kappa_2)_{\max}-(\cos\kappa_2)_{\min}
 =\frac{2|\mathcal B K_1|}{\sqrt{\mathcal A^2+\mathcal B^2}}
 \leq\frac{2|\mathcal B|}{\sqrt{\mathcal A^2+\mathcal B^2}}\ ,
 \label{Eq weakspin-amplitude}
\end{equation}
since $|K_1|\leq 1$. Thus, a variation approaching the full interval from $+1$ to $-1$ is possible~when
\begin{equation}
 |\mathcal A|\ll|\mathcal B|\ .
 \label{Eq large-ff-condition}
\end{equation}
This condition allows a large flip-flop (defined as a polar-angle excursion that is a substantial fraction of the full $0\leq\kappa_i\leq\pi$ range; the limiting case $\Delta\kappa_i=\pi$ corresponds to an~aligned-to-anti-aligned sweep) but does not guarantee one: the initial angles must also give a sufficiently large $|K_1|$ and place the mean value $K_2$ sufficiently close to zero. These conditions depend on the initial values of $\kappa_2$ and $\Delta \zeta$. The equations must permit a large oscillation, and the system must begin on a trajectory that uses most of the permitted range. Near equal masses, Equation~\eqref{Eq large-ff-condition} can be satisfied in two ways. For a Kerr-dominant object, $w_1=1$, the usual large flip-flop occurs when
\begin{equation}
 \cos\kappa_1=\mathcal{O}(v)\ ,
 \label{Eq standard-ff-condition}
\end{equation}
so the dominant spin lies close to the orbital plane, while $\mathcal B$ remains large because $\kappa_1\approx\pi/2$ and $\mathcal{B}\propto x_1 \sin\kappa_1$. The quadrupole-induced case instead follows from
\begin{equation}
 1+2q^{-1}-w_1-q^{-1}w_1x_1\cos\kappa_1=\mathcal{O}(v).
 \label{Eq quadrupole-ff-condition}
\end{equation}
Since $x_1=\mathcal{O}(v)$, the leading-order condition results in
\begin{equation}
 w_1\simeq1+\frac{2}{q}\ ,
\end{equation}
and hence $w_1\simeq3$ for $q\simeq1$. This is the quadrupole-induced flip-flop identified in Ref.~\cite{KeresztesTapaiGergely2021}. When gravitational radiation is taken into account, the harmonic solution should no longer assume that $v, \kappa, \mathcal{A}$, and $\mathcal{B}$ are constant during a cycle. In a realistic inspiral, these parameters change slowly. The motion is therefore only approximately periodic; indeed, successive flip-flops can have different periods and amplitudes.

\subsection{Adiabatic 2PN Approximation}
Gravitational radiation removes energy from the binary and causes the orbital frequency to increase. In the adiabatic approximation, the inspiral is treated as a sequence of nearly circular orbits. The orbital energy $E(v)$ and the gravitational-wave energy flux are related by $dE/dt=-\mathcal{F}_{\rm GW}$. Expanding the circular-orbit energy and flux using Equation~\eqref{Eq v and tau} through 2PN order gives~\cite{MikocziVasuthGergely2005,Kidder1995}
\begingroup\makeatletter\def\f@size{9}\check@mathfonts
\def\maketag@@@#1{\hbox{\m@th\normalsize \normalfont#1}}%
\begin{equation}
 \frac{dv}{d\tau}=\frac{32}{5}\eta v^9
 \left[1-\left(\frac{743}{336}+\frac{11}{4}\eta\right)v^2
 +(4\pi-\beta)v^3
 +\left(\frac{34103}{18144}+\frac{13661}{2016}\eta
 +\frac{59}{18}\eta^2+\sigma\right)v^4\right].
 \label{Eq flux}
\end{equation}\endgroup
The leading factor, $\frac{32}{5}\eta v^9$, is the Newtonian result. Within the square brackets, the terms proportional to $v^2$, $v^3$, and $v^4$ are the relative 1PN, 1.5PN, and 2PN corrections, respectively; there is no 0.5PN term for a quasi-circular binary. At 1.5PN order, $4\pi v^3$ is the nonspinning gravitational-wave tail term arising from radiation interacting with the curved spacetime generated by the binary itself. The other 1.5PN contribution is $-\beta v^3$, the leading spin--orbit term, where
\begin{equation}
 \beta=\frac{1}{12}\sum_{i=1}^{2}\chi_i\cos\kappa_i
 \left[113\left(\frac{m_i}{M}\right)^2+75\eta\right]\ .
 \label{Eq beta}
\end{equation}
Thus, each spin first enters the chirp through its signed projection $\chi_i\cos\kappa_i$ along $\bm L_N$. When spin $i$ crosses the orbital plane, $\cos\kappa_i$ changes sign and so does that body's contribution to $\beta$, while the spin magnitude $\chi_i$ remains constant. This explains why a visually large inversion of a weak spin may have little effect on the waveform. Notice that the total $\beta$ does not necessarily change sign when one spin flips. The other spin may still dominate the sum.

The leading terms that are quadratic in the spins enter at 2PN order through
\begin{equation}
 \sigma=\sigma_{S_1S_2}+\sigma_{\mathrm{self}}+\sigma_{\mathrm{QM}},
 \label{Eq sigma}
\end{equation}
with\vspace{-12pt}
\begin{align}
 \sigma_{S_1S_2}&=\frac{\eta\chi_1\chi_2}{48}
 \left(721\cos\kappa_1\cos\kappa_2-247\cos\gamma\right),\\
 \sigma_{\mathrm{self}}&=\frac{1}{96}\sum_{i=1}^{2}
 \chi_i^2\left(\frac{m_i}{M}\right)^2\left(6+\sin^2\kappa_i\right),\\
 \sigma_{\mathrm{QM}}&=\frac{5}{2}\sum_{i=1}^{2}w_i\chi_i^2
 \left(\frac{m_i}{M}\right)^2\left(3\cos^2\kappa_i-1\right).
\end{align}
The mutual spin--spin term $\sigma_{S_1S_2}$ depends on both spin projections and on the angle $\gamma$ between the spins. The angle $\gamma$ is related to the three internal angles by
Equation~\eqref{Eq gamma}. The self-spin term $\sigma_{\text{self}}$ is associated with each spin separately. Because it contains $\sin^2\kappa$, it does not distinguish directly between corresponding aligned-side and anti-aligned-side orientations. The quadrupole--monopole term $\sigma_{\text{QM}}$ arises because the spinning compact object has a mass quadrupole. The parameter $w_i$ specifies the quadrupole produced by a given spin magnitude
through Equation~\eqref{Eq spin-quadrupole}.
Because $\sigma_{\text{QM}}$ depends on $\cos^2\kappa_i$, it  cannot tell whether the spin projection is positive or negative at two mirror-related orientations; it is not directly sensitive to the sign change at an exact orbital-plane crossing. The magnetic-dipole term in the general expression from Ref.~\cite{MikocziVasuthGergely2005} is zero for the systems considered here and is omitted.

\subsection{Restricted Waveform Model and LISA Weighting}
\label{2.4}
We construct a waveform model for the dominant quadrupole harmonic of a quasi-circular inspiral. It is a restricted numerical waveform constructed using the stationary phase approximation (SPA), with a Newtonian quadrupole source amplitude and a TaylorT4-type 2PN chirp, coupled to the secular spin-angle equations. Its frequency-domain carrier is obtained by applying the SPA to a numerical evolution of the orbital frequency and phase, coupled to the secular spin-angle equations. The source amplitude is kept at Newtonian order. The orbital-frequency evolution is given by Equation~\eqref{Eq flux}, including the nonspinning terms through relative 2PN order, the leading spin--orbit term at 1.5PN order, and the mutual spin--spin, self-spin, and quadrupole--monopole terms at 2PN order~\cite{MikocziVasuthGergely2005,GergelyMikoczi2009,Kidder1995}. The three internal spin angles obey Equations~\eqref{Eq kappa derivatives1}--\eqref{Eq deltazeta derivative}. These five evolution equations supply the frequency-dependent spin geometry and the gravitational-wave phase used below. The gravitational wave phase satisfies
\begin{equation}
 \frac{d\phi_{\rm GW}}{d\tau}=2v^3,
 \label{Eq gw phase derivative}
\end{equation}
because $\phi_{\rm GW}=2\phi_{\rm orb}$ and $d\phi_{\rm orb}/dt=\Omega=v^3/M$. Since $f=v^3/(\pi M)$, the frequency derivative required by the SPA is
\begin{equation}
 \dot f\equiv\frac{df}{dt}
 =\frac{3v^2}{\pi M^2}\frac{dv}{d\tau},
 \label{Eq frequency derivative}
\end{equation}
where $dv/d\tau$ is evaluated from Equation~\eqref{Eq flux}. At every sampled value of $v$, the evolved variables
$c_1$, $c_2$, and $c_\gamma$ determine $\beta$ and $\sigma$. The angle evolution therefore changes both the accumulated phase and the instantaneous chirp rate.

We use the Fourier convention $\widetilde h(f)=\int h(t)e^{2\pi i f t}dt$. For $f>0$, the SPA gives
\begin{equation}
 \widetilde h(f)=A(f)e^{i\Psi(f)}\ ,\qquad
 \Psi(f)=2\pi f t_f-\phi_{\rm GW}(t_f)-\frac{\pi}{4}\ ,
 \label{eq:spa-waveform}
\end{equation}
where the stationary time $t_f$ is defined by $d\phi_{\rm GW}/dt|_{t_f}=2\pi f$. Thus, $v(t_f)=(\pi Mf)^{1/3}$, and the numerical solution directly supplies $t(f)$ and $\phi_{\rm GW}(f)$. Equation~\eqref{eq:spa-waveform} is not an explicit TaylorF2 polynomial with constant spin projections. It is obtained by integrating the 2PN-truncated chirp equation while recalculating $\beta$ and $\sigma$ from the evolving spin angles, and then applying the SPA. The numerical integration of this truncated differential equation is not algebraically identical to an explicit TaylorF2 re-expansion, although the two have the same formal terms through 2PN when the spin angles are fixed. No nonspinning terms above 2PN, spin--orbit terms above 1.5PN, quadratic-spin terms above 2PN, or cubic and higher powers of the spins are included. For the amplitude, we define
\begin{equation}
 \mathcal Q(\iota)=
 \left[\left(\frac{1+\cos^2\iota}{2}\right)^2
 +\cos^2\iota\right]^{1/2}\ ,
 \label{Eq polarization amplitude factor}
\end{equation}
where the two terms are the squared Newtonian inclination factors of the plus and cross polarizations. We combine them in quadrature rather than applying detector antenna patterns. With luminosity distance $D_L$, the implemented positive-frequency amplitude is
\begin{equation}
 A(f)=\frac{2\eta M}{D_L}\,
 \frac{v^2}{\sqrt{\dot f}}\,\mathcal{Q}[\iota(f)]\ ,
 \qquad v=(\pi Mf)^{1/3}\ .
 \label{Eq restricted amplitude}
\end{equation}
If only the Newtonian chirp rate
\begin{equation}
 \dot f_{\rm N}=\frac{96}{5\pi M^2}\eta v^{11}
\end{equation}
is inserted into Equation~\eqref{Eq restricted amplitude}, then
\begin{equation}
 A_{\mathrm{N}}(f)=\sqrt{\frac{5}{24}}\,
 \frac{M_{\rm c}^{5/6}}{\pi^{2/3}D_L}\,
 f^{-7/6}\mathcal Q[\iota(f)]\ ,
 \qquad M_{\rm c}=\eta^{3/5}M\ .
 \label{Eq Newtonian frequency amplitude}
\end{equation}
In the calculation, however, $\dot f$ in Equation~\eqref{Eq restricted amplitude} is obtained from the full 2PN expression in Equation~\eqref{Eq flux}. Consequently, the SPA Jacobian $1/\sqrt{\dot f}$ contains the PN corrections implied by the 2PN chirp, including its evolving spin terms.

It remains to specify the inclination $\iota(f)$ appearing in
Equations~\eqref{Eq polarization amplitude factor} and
\eqref{Eq restricted amplitude}. We approximate
$\widehat{\bm J}(t)\simeq\widehat{\bm J}(t_0)$ during the modeled
inspiral. Let $\alpha_L$ be the angle between $\bm L_N$ and $\bm J$,
$\theta_{JN}$ the angle between $\bm J$ and the line of sight $\bm N$,
and $\lambda_L$ the azimuth of $\bm L_N$ about this fixed $\bm J$ direction.
For $\theta_{JN}=0^\circ$, the viewing direction is along the total angular
momentum and $\iota=\alpha_L$. Observer-frame precession signatures can be stronger for generic
viewing directions~\cite{FairhurstEtAl2020}. Using
\begin{equation}
 \bm J\simeq\bm L_N+\bm S_1+\bm S_2,
 \qquad L_N=\frac{\eta M^2}{v},
\end{equation}
where $L_N$ is evaluated at Newtonian order, gives
\begin{equation}
 \cos\alpha_L=
 \frac{L_N+S_1\cos\kappa_1+S_2\cos\kappa_2}
 {\left[
 L_N^2+S_1^2+S_2^2
 +2L_N(S_1\cos\kappa_1+S_2\cos\kappa_2)
 +2S_1S_2\cos\gamma
 \right]^{1/2}}.
 \label{eq:inclination-factor}
\end{equation}
For the inclination extension, we reconstruct $\lambda_L$ by integrating
the circular limit of the secular nodal rate in Equations~(15)--(17) from
Ref.~\cite{KeresztesTapaiGergely2021}. In that nodal convention,
$d\lambda_L/d\tau=-d\phi_n/d\tau$, with $\tau=t/M$. The spin--azimuth
combinations entering the rate are reconstructed algebraically from the
internal spin angles and the total-angular-momentum constraint. Taking the
observer azimuth to be zero gives
\begin{equation}
 \cos\iota(f)=\cos\alpha_L(f)\cos\theta_{JN}
 +\sin\alpha_L(f)\sin\theta_{JN}\cos\lambda_L(f).
 \label{eq:generic-viewing-inclination}
\end{equation}
The injected initial orbital azimuth is $\lambda_{L,0}=0$.
For each nonzero $\theta_{JN}$, the template initial azimuth is varied over
$[0,2\pi)$; at $\theta_{JN}=0$, it is unobservable and fixed to zero.
The template and injection have the same fixed $\theta_{JN}$ in each
comparison. The extension changes only the quadrature inclination factor $\mathcal Q$.
Thus, the leading amplitude changes through both the evolving orbital cone
and its azimuthal projection onto the observer direction.

Within these limits, spin inversions enter the waveform through $\beta(f)$ and $\sigma(f)$ in the accumulated phase, through the same coefficients in the factor $1/\sqrt{\dot f}$, and through $\iota(f)$ in $\mathcal Q$. The reported mismatches include the accumulated phase change and the leading inclination-amplitude change. The overall distance factor cancels from normalized matches. We use the analytic sky-averaged LISA sensitivity, including the one-year Galactic confusion foreground, from Ref.~\cite{RobsonCornishLiu2019}. This supplies the frequency weighting but not a moving-detector response~\cite{LISA2017}. The adopted calculation assumes uninterrupted observation over the selected frequency interval and a stationary effective noise PSD. Data gaps, non-stationary instrumental artifacts, overlapping sources, and other contamination are not modeled here and can further reduce practical distinguishability.

For a signal $h$, we define the noise-weighted squared signal amplitude as
\begin{equation}
 (h|h)=4\int_{f_{\min}}^{f_{\max}}
 \frac{|\widetilde h(f)|^2}{S_n(f)}\,df\ ,
\end{equation}
For stationary Gaussian noise, $(h|h)$ is the squared noise-weighted norm of the signal and equals the squared optimal matched-filter SNR for a perfectly matched template; equivalently, $\rho[h]=\sqrt{(h|h)}$. For two signals, the complex correlation at a relative time shift $\Delta t$ is
\begin{equation}
 z(\Delta t)=4\int_{f_{\min}}^{f_{\max}}
 \frac{\widetilde h_1(f)\widetilde h_2^*(f)}{S_n(f)}
 e^{2\pi i f\Delta t}\,df \ .
\end{equation}
Their match is
\begin{equation}
 \mathcal M(h_1,h_2)=\max_{\Delta t}
 \frac{|z(\Delta t)|}
 {\sqrt{(h_1|h_1)(h_2|h_2)}} .
 \label{eq:waveform-match}
\end{equation}
Taking the absolute value of $z$ maximizes the correlation over a
constant phase difference. We evaluate the frequency integral by
trapezoidal quadrature for a grid of time shifts. After locating the
largest sampled value of $|z(\Delta t)|$, a continuous one-dimensional
maximization refines the time shift in the neighboring interval. The
reported match is therefore not restricted to the discrete time-shift
grid.

To determine whether a spin-inverting signal can be reproduced by a system without an inversion, we compare it with waveforms whose spin angles continue to evolve but for which neither $\cos\kappa_1$ nor $\cos\kappa_2$ reaches or crosses zero. We perform separate searches in the four possible sign combinations
\begin{equation}
 (+,+)\ \qquad (+,-)\ \qquad (-,+)\ \qquad (-,-)
\end{equation}
where the first sign is the sign of $\cos\kappa_1$ and the second is the sign of $\cos\kappa_2$ throughout the analyzed frequency interval. For example, the $(-,+)$ case requires $\cos\kappa_1<0$ and $\cos\kappa_2>0$ at every integration point. Any candidate that reaches or crosses zero is rejected. Within each sign combination, we vary the mass ratio, spin magnitudes, initial spin angles, and total mass. If $\Theta_{\rm NI}$ denotes the set of all varied parameters that satisfy the no-inversion condition, the largest match found is
\begin{equation}
 \mathcal M_{\rm best}
 =\max_{\bm\theta\in\Theta_{\rm NI}}
 \mathcal M\!\left[h_{\rm inv},h_{\rm NI}(\bm\theta)\right]\ .
 \label{eq:best-no-inversion-match}
\end{equation}
Here, $h_{\rm inv}$ is the spin-inverting injection and
$h_{\rm NI}(\bm\theta)$ is a physically evolving no-inversion waveform. Throughout, ``best no-inversion waveform'' means the largest-match no-inversion candidate found within the restricted GR waveform model, with the quadrupole prescription and parameter ranges specified here and in Appendix~\ref{app:no-inversion-search}. Specific astrophysical environments or more accurate source parametrizations would require their own consistent no-inversion waveforms. Because the match divides by the amplitudes of both signals, it is unchanged by an overall constant rescaling of either waveform. This is equivalent to fitting one constant amplitude factor, in addition to the time and phase shifts, when the residual is calculated.

If the spin-inverting signal is assigned a reference SNR $\rho_\star$, the SNR remaining after the subtraction of the best amplitude-fitted no-inversion waveform is
\begin{equation}
 \rho_{\rm res}
 =\rho_\star\sqrt{1-\mathcal M_{\rm best}^2}.
 \label{eq:best-residual-snr}
\end{equation}
For a small best-fit mismatch
$\epsilon_{\rm best}=1-\mathcal M_{\rm best}$, this reduces to $\rho_{\rm res}
 \simeq\rho_\star\sqrt{2\epsilon_{\rm best}}\ $~\cite{LindblomOwenBrown2008}. We use $\rho_\star=100$ when comparing the injections. These calculations
measure how well the tested no-inversion waveforms reproduce a spin-inverting
signal within the restricted waveform model. They do not replace Bayesian
parameter estimation with a complete precessing waveform and the full LISA
response. Complete precessing Kerr-binary waveforms, including IMRPhenomXPHM and SEOBNRv5PHM, perform an inertial-frame rotation and include merger, ringdown, and higher modes~\cite{PrattenEtAl2021,RamosBuadesEtAl2023}. Poor matches among different precessing gravitational-wave models have also been investigated~\cite{TapaiEtAl2018}. A modification of IMRPhenomXPHM also includes a spin-induced quadrupole in the phase and precession dynamics~\cite{LyuEtAl2024}. The broader waveform requirements for LISA, including accuracy, computational efficiency, physical completeness, and detector-response modeling, have recently been reviewed by the LISA Consortium Waveform Working Group~\cite{LISAConsortium2025}. At the PN inspiral level, the recent pyEFPEHM model combines eccentricity, spin precession, higher-order modes, and matter effects within a single waveform framework~\cite{Morras2026}. We do not alter the precession equations of these models here. 

\section{Observational Distinguishability of Spin Inversions}

\subsection{Spin Evolution}
\label{subsec:aligned-view-recovery}

\begin{figure}[H]
\includegraphics[scale=0.69]{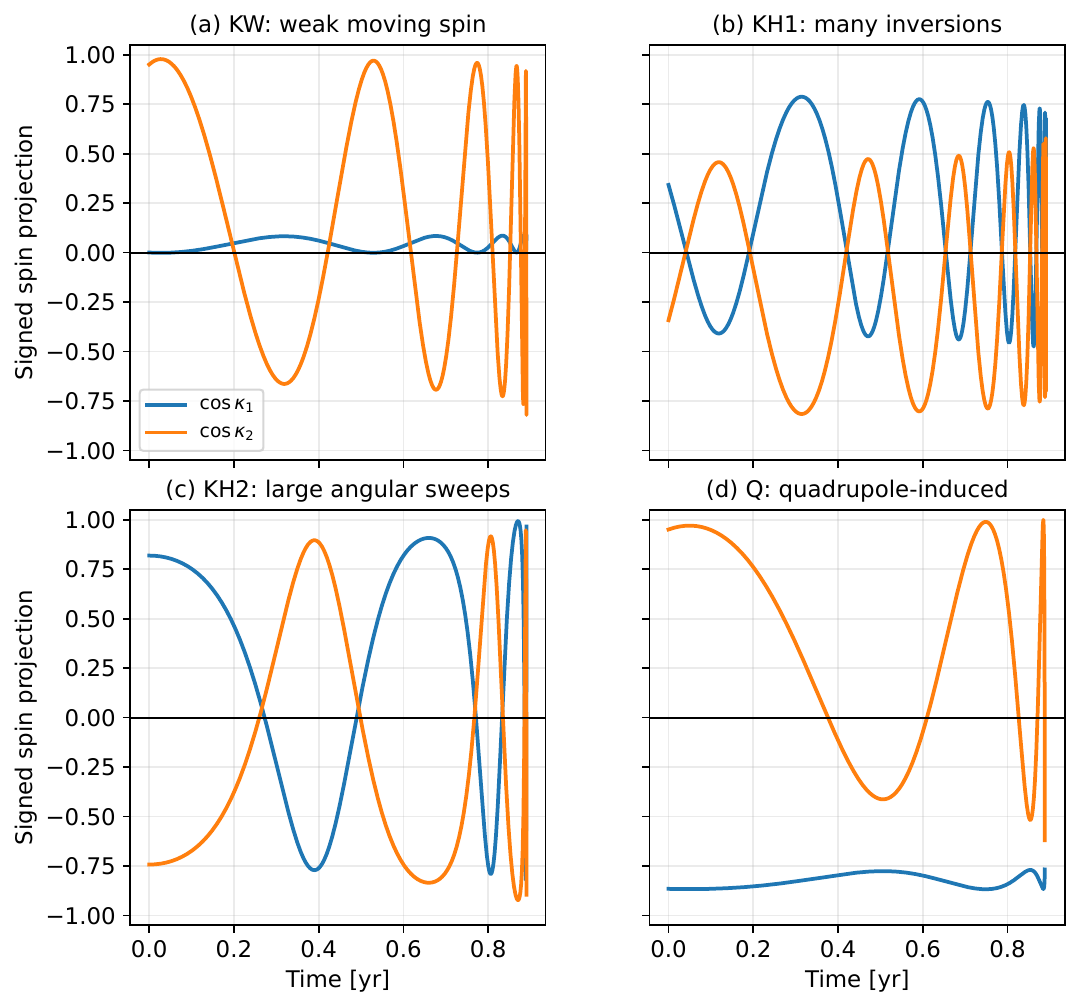}
\caption{Spin dynamics of four of the selected injections listed in Table%
~\ref{tab:injection-parameters} at $\theta_{JN}=0^\circ$.
The curves show $\cos\kappa_i=\widehat{\bm S}_i\!\cdot\!\widehat{\bm L}_N$.
A zero crossing is an inversion of spin $i$ relative to the orbital plane,
while repeated alternating crossings indicate flip-flop motion.  Panels (\textbf{a}--\textbf{c}) show Kerr configurations with, respectively,
a weak moving secondary, many inversions of two large spins, and large angular
sweeps of two large spins.  Panel (\textbf{d}) shows the quadrupole-induced Q
configuration with $w_1=3.0619$ and $w_2=3$.}
\label{fig:spin-dynamics}
\end{figure}

We now ask whether the spin inversions produced by the secular equations leave
a waveform difference that cannot be reproduced by a binary without an
orbital-plane crossing. We use five injections. Four have Kerr quadrupole
coefficients, $w_1=w_2=1$, and sample weak-spin and two-active-spin motion. The
fifth is a non-Kerr configuration chosen from the quadrupole-induced case, with $w_1=1+2/q=3.0619$ and $w_2=3$. All cases include a total mass of
$M=2\times10^5M_\odot$, a luminosity distance of $D_L=1000\,{\rm Mpc}$ and are evolved over
$2\times10^{-4}\,{\rm Hz}\leq f\leq10^{-2}\,{\rm Hz}$. This frequency interval corresponds to
approximately $0.085\leq v\leq0.314$. The injection parameters are given in
Table~\ref{tab:injection-parameters}. All masses are detector-frame masses.
The luminosity distance fixes the waveform amplitude before normalization, but
does not affect the matches reported below. Four cases of the signed projections $\cos\kappa_i$ are shown in
Figure~\ref{fig:spin-dynamics}. The case labels summarize the purpose of each injection. K0 is the
baseline Kerr configuration with two substantial spins, while KW uses
the same initial geometry with a weak secondary spin. KH1 and KH2 are
high-spin Kerr configurations selected to exhibit many inversions and
large angular excursions, respectively; Q denotes the non-Kerr,
quadrupole-induced configuration.
\vspace{-5pt}

\begin{table}[H]
\centering
\caption{Parameters of the spin-inverting injections. The angles are their
values at the initial frequency.}
\label{tab:injection-parameters}

\begin{adjustwidth}{-\extralength}{0cm}
\begin{tabularx}{\fulllength}{LCCCCCCCC}
\toprule
\textbf{Case} & \boldmath{$q$} & \boldmath{$\chi_1$} & \boldmath{$\chi_2$}
& \boldmath{$\kappa_{1,0}$} & \boldmath{$\kappa_{2,0}$} & \boldmath{$\Delta\zeta_0$}
& \boldmath{$w_1$} & \boldmath{$w_2$} \\
& & & & \textbf{(deg)} & \textbf{(deg)} & \textbf{(deg)} & & \\
\midrule
K0  & 0.97 & 0.95 & 0.50 & 90  & 18  & 135 & 1      & 1 \\
KW  & 0.97 & 0.95 & 0.05 & 90  & 18  & 135 & 1      & 1 \\
KH1 & 0.94 & 0.95 & 0.95 & 70  & 110 & 80  & 1      & 1 \\
KH2 & 0.97 & 0.95 & 0.95 & 35  & 138 & 180 & 1      & 1 \\
Q   & 0.97 & 0.70 & 0.05 & 150 & 18  & 45  & 3.0619 & 3 \\
\bottomrule
\end{tabularx}
\end{adjustwidth}
\end{table}

\vspace{-5pt}
The KW system demonstrates that a spin with small
magnitude can move through a large angle: the secondary has $\chi_2=0.05$ but
spans $133^\circ$ and crosses the plane nine times in the calculated
frequency interval.  Increasing the secondary spin to $\chi_2=0.50$ in K0
produces a $155^\circ$ secondary-spin excursion, while the primary moves
through $80^\circ$.  In KH1 and KH2, both spins take part in the motion.  KH1
has fourteen crossings of each spin, whereas KH2 has fewer crossings but
larger angular excursions.  In the Q configuration, the dominant spin remains
anti-aligned and moves by only $10^\circ$, while the weak secondary spans
$128^\circ$ and crosses the orbital plane five times.

Table~\ref{tab:spin-dynamics} gives the corresponding durations, angular
excursions, and crossing counts.  The number of crossings and the total angular
excursion measure different properties.  In particular, K0 and KW start with
$\kappa_{1,0}=90^\circ$, so some primary-spin crossings are shallow passages
near the orbital plane.  They should not be interpreted as complete
$180^\circ$ primary-spin flip-flops.  The maximum error in the angular
constraint is $1.6\times10^{-9}$ among the five~integrations.

\vspace{-3pt}
\begin{table}[H]
\centering
\caption{Spin motion of the injections.  Here, $N_i$ is the number of
sign changes in $\cos\kappa_i$ over the calculated frequency interval.}
\label{tab:spin-dynamics}
\begin{tabularx}{\textwidth}{L c CC CC}
\toprule
\textbf{Case} & \textbf{Duration (yr)} & \boldmath{$\Delta\kappa_1$} \textbf{(deg)} & \boldmath{$\Delta\kappa_2$} \textbf{(deg)} & \boldmath{$N_1$} & \boldmath{$N_2$} \\
\midrule
K0  & 0.8919 & 79.56  & 154.75 & 3  & 6  \\
KW  & 0.8904 & 5.27   & 133.26 & 5  & 9  \\
KH1 & 0.8911 & 83.56  & 90.09  & 14 & 14 \\
KH2 & 0.8906 & 138.78 & 138.81 & 6  & 6  \\
Q   & 0.8881 & 10.10  & 128.11 & 0  & 5  \\
\bottomrule
\end{tabularx}
\end{table}
\vspace{-5pt}
The recovery comparison in this subsection uses the aligned viewing
direction $\theta_{JN}=0$. A large angular excursion does not by itself
establish that the inversion can be distinguished in detector data.  The relevant comparison is with the
physical no-inversion waveform that gives the largest match to the injection.
For each case, we searched the four sectors
$(+,+)$, $(+,-)$, $(-,+)$, and $(-,-)$, where the signs specify the required
signs of $(\cos\kappa_1,\cos\kappa_2)$.  Every trial waveform was evolved with
the same secular equations as the injection.  It was retained only if neither
spin crossed the orbital plane.  The detector-frame total mass, mass ratio,
spin magnitudes, and three initial spin angles were allowed to vary; the
quadrupole coefficients were fixed to those of the corresponding injection.
The match was maximized over a constant time shift and phase shift, and an
overall amplitude factor was fitted when the residual was calculated.  We report the largest match found and the corresponding
smallest residual found. We use the quantity
\begin{equation}
 \mu_i=\min_f\left[s_i\cos\kappa_i(f)\right],
 \label{eq:no-inversion-margin}
\end{equation}
where $s_i=+1$ or $-1$ is fixed by the selected sector. A positive value of $\mu_i$ verifies that spin $i$ remains in the prescribed no-inversion sector. All fitted waveforms in Table~\ref{tab:no-inversion-results} satisfy this
condition and have zero detected crossings on the numerical frequency grid.
The complete search intervals, optimization settings, best-fit
no-inversion parameters, and numerical verification tests are reported in
Appendix~\ref{app:no-inversion-search}.

\vspace{-3pt}
\begin{table}[H]
\centering
\caption{No-inversion matches and residual SNRs after rescaling each injection to
$\rho_\star=100$ at $\theta_{JN}=0^\circ$.}
\label{tab:no-inversion-results}
\small
\begin{tabularx}{\textwidth}{l C CC C c}
\toprule
\textbf{Case} & \textbf{Sector} & \boldmath{$\mathcal M_{\rm best}$}
& \boldmath{$1-\mathcal M_{\rm best}$}
& \boldmath{$\rho_{\rm res}$}
& \boldmath{$(\mu_1,\mu_2)$} \\
\midrule
K0  & $(+,+)$ & 0.9999720034 & $2.800\times10^{-5}$ & 0.748 & $(0.069,\,0.521)$ \\
KW  & $(+,+)$ & 0.9999674995 & $3.250\times10^{-5}$ & 0.806 & $(0.044,\,0.024)$ \\
KH1 & $(-,+)$ & 0.9998652957 & $1.347\times10^{-4}$ & 1.641 & $(0.230,\,0.185)$ \\
KH2 & $(+,+)$ & 0.9999731354 & $2.686\times10^{-5}$ & 0.733 & $(0.655,\,0.716)$ \\
Q   & $(-,-)$ & 0.9999897071 & $1.029\times10^{-5}$ & 0.454 & $(0.206,\,0.719)$ \\
\bottomrule
\end{tabularx}%
\end{table}
\vspace{-5pt}
For each injection and its selected no-inversion waveform in Table~\ref{tab:no-inversion-results}, the signal SNR at which $\rho_{\rm res}=1$ is $\rho_{\star,1}=1/\sqrt{1-\mathcal M_{\rm best}^2}$. The corresponding values are approximately 134, 124, 61, 136, and 220 for K0, KW, KH1, KH2, and Q, respectively~\cite{DamourNagarTrias2011}.

The aligned-view recovery results are listed in
Table~\ref{tab:no-inversion-results}. At $\rho_\star=100$, the largest
residual is $\rho_{\rm res}=1.64$ for KH1.  The other four residuals are
smaller than unity.  KW therefore supplies a direct example in which a
$133^\circ$ motion of the weak spin is almost completely reproduced by a
no-inversion source.  K0 and KH2 show that increasing the spin magnitude or the
angular excursion does not necessarily increase the remaining waveform
difference.  Even KH1, which contains fourteen crossings of each spin, is
closely matched after the source parameters are varied.

For the aligned-view fits, we also tested whether the full-band overlap hides a more distinctive difference during an inversion. For this, we defined for each zero crossing of $\cos\kappa_i(t)$, an individual inversion window bounded by the neighboring turning points of the same spin projection, so that it contains one monotonic passage through the orbital plane. Within the SPA used here, the monotonic time-frequency relation $t(f)$ maps each such time interval to the corresponding frequency range. We retained the intrinsic parameters of the best full-band no-inversion candidate, while re-fitting only the constant time shift, phase shift, and amplitude in the selected interval. The local windows were not renormalized, the injection remaining normalized to the full-band SNR value $\rho_\star=100$.
We first considered a single interval containing all inversions, from the turning point preceding the first crossing to the turning point following the last crossing. The residual SNRs are $0.75$, $0.63$, $1.35$, $0.73$, and $0.45$ for K0, KW, KH1, KH2, and Q, respectively, compared with the corresponding full-band values $0.75$, $0.81$, $1.64$, $0.73$, and $0.45$. We then examined the individual inversion-containing windows separately. The largest residual SNRs among these windows are $0.59$, $0.18$, $0.32$, $0.29$, and $0.27$, respectively, again all below the full-band values.
The normalized mismatch can nevertheless increase locally. For KH2, it rises from $2.69\times10^{-5}$ over the full band to about $5.3\times10^{-5}$ in the most discrepant individual window, while for Q it rises from $1.03\times10^{-5}$ to about $6.9\times10^{-5}$. The increase in normalized mismatch does not imply a larger absolute residual, because these short windows contain only a small fraction of the total signal power. Their signal SNRs are only $\rho_W\simeq1.62$ and $1.22$, respectively. With $\rho_{\rm res,W}=\rho_W\sqrt{1-\mathcal M_W^2}$, the corresponding residual SNRs are therefore only $0.017$ and $0.014$. Thus, localization can make the fractional waveform difference more pronounced without increasing the absolute residual SNR at fixed physical signal strength.

\subsection{Dependence on Observer Inclination}
\label{subsec:observer-inclination}

Precession is difficult to observe when the total angular momentum is close to the line of sight, whereas its imprint can be appreciably stronger for generic viewing directions~\cite{FairhurstEtAl2020}. We therefore performed the four-sector no-inversion comparison for all five injections at $\theta_{JN}=0^\circ,15^\circ,30^\circ,45^\circ,50^\circ,55^\circ,60^\circ,65^\circ,70^\circ,75^\circ$, and $90^\circ$, using Equation~\eqref{eq:generic-viewing-inclination}. To describe these directions consistently, we supplemented the internal spin-angle evolution by reconstructing the orbital azimuth from angular-momentum conservation, as detailed in Section~\ref{2.4}. The intrinsic template parameters were searched separately for each case and inclination, and the initial orbital azimuth was also varied at nonzero inclinations.
The consolidated selection contains $55$ fits, evaluated at $8192$ frequency
samples. The zero-degree entries come from independent searches and subsequent
refinement, and are distinct from the fit set in
Table~\ref{tab:no-inversion-results}. Their residual SNRs at $\rho_\star=100$
are $0.217$, $0.806$, $1.608$, $0.356$, and $0.446$ for K0, KW, KH1,
KH2, and Q, respectively.

Each injection is separately normalized to $\rho_\star=100$ at each viewing direction, so the comparison concerns waveform shape at equal injected SNR rather than inclination dependence at fixed physical distance. KW remains below residual SNR unity throughout, whereas the other cases reach larger values at several nonzero inclinations. Table~\ref{tab:inclination-results} gives the residual SNR and mismatch for every case and sampled inclination; the largest tabulated value is $7.068$ for KH1 at $55^\circ$.

\begin{table}[H]
\centering
\caption{Residual %
 SNR and mismatch of the selected no-inversion fit for all
five cases at the eleven sampled observer inclinations, evaluated at $8192$
frequency samples and normalized to $\rho_\star=100$.}
\label{tab:inclination-results}
\small
\begin{adjustwidth}{-\extralength}{0cm}
\centering
{%
\begin{tabularx}{\fulllength}{C ccccc ccccc c}
\toprule
& \multicolumn{2}{c}{\textbf{K0}} & \multicolumn{2}{c}{\textbf{KW}}
& \multicolumn{2}{c}{\textbf{KH1}} & \multicolumn{2}{c}{\textbf{KH2}}
& \multicolumn{2}{c}{\textbf{Q}} \\
\cmidrule{2-11}
\boldmath{$\theta_{JN}$} \textbf{(deg)}
& \boldmath{$\rho_{\rm res}$} & \boldmath{$\epsilon_{\rm best}$}
& \boldmath{$\rho_{\rm res}$} & \boldmath{$\epsilon_{\rm best}$}
& \boldmath{$\rho_{\rm res}$} & \boldmath{$\epsilon_{\rm best}$}
& \boldmath{$\rho_{\rm res}$} & \boldmath{$\epsilon_{\rm best}$}
& \boldmath{$\rho_{\rm res}$} & \boldmath{$\epsilon_{\rm best}$} \\
\midrule
0 & 0.217 & $2.347 \times 10^{-6}$ & 0.806 & $3.250 \times 10^{-5}$ & 1.608 & $1.292 \times 10^{-4}$ & 0.356 & $6.339 \times 10^{-6}$ & 0.446 & $9.929 \times 10^{-6}$ \\
15 & 1.585 & $1.256 \times 10^{-4}$ & 0.919 & $4.218 \times 10^{-5}$ & 3.977 & $7.913 \times 10^{-4}$ & 2.101 & $2.208 \times 10^{-4}$ & 1.505 & $1.132 \times 10^{-4}$ \\
30 & 2.143 & $2.297 \times 10^{-4}$ & 0.897 & $4.024 \times 10^{-5}$ & 4.204 & $8.842 \times 10^{-4}$ & 2.195 & $2.410 \times 10^{-4}$ & 2.514 & $3.159 \times 10^{-4}$ \\
45 & 2.447 & $2.994 \times 10^{-4}$ & 0.910 & $4.141 \times 10^{-5}$ & 6.351 & $2.019 \times 10^{-3}$ & 4.151 & $8.620 \times 10^{-4}$ & 2.299 & $2.644 \times 10^{-4}$ \\
50 & 4.553 & $1.037 \times 10^{-3}$ & 0.915 & $4.184 \times 10^{-5}$ & 5.343 & $1.428 \times 10^{-3}$ & 5.334 & $1.424 \times 10^{-3}$ & 1.875 & $1.758 \times 10^{-4}$ \\
55 & 3.292 & $5.420 \times 10^{-4}$ & 0.918 & $4.215 \times 10^{-5}$ & 7.068 & $2.501 \times 10^{-3}$ & 5.711 & $1.632 \times 10^{-3}$ & 2.523 & $3.184 \times 10^{-4}$ \\
60 & 3.274 & $5.361 \times 10^{-4}$ & 0.919 & $4.222 \times 10^{-5}$ & 6.793 & $2.310 \times 10^{-3}$ & 4.119 & $8.489 \times 10^{-4}$ & 1.567 & $1.228 \times 10^{-4}$ \\
65 & 2.745 & $3.769 \times 10^{-4}$ & 0.915 & $4.185 \times 10^{-5}$ & 6.641 & $2.208 \times 10^{-3}$ & 2.298 & $2.640 \times 10^{-4}$ & 3.827 & $7.325 \times 10^{-4}$ \\
70 & 1.735 & $1.506 \times 10^{-4}$ & 0.904 & $4.089 \times 10^{-5}$ & 6.491 & $2.109 \times 10^{-3}$ & 4.440 & $9.862 \times 10^{-4}$ & 3.964 & $7.861 \times 10^{-4}$ \\
75 & 2.870 & $4.120 \times 10^{-4}$ & 0.888 & $3.943 \times 10^{-5}$ & 6.212 & $1.931 \times 10^{-3}$ & 5.944 & $1.768 \times 10^{-3}$ & 3.241 & $5.252 \times 10^{-4}$ \\
90 & 1.492 & $1.114 \times 10^{-4}$ & 0.864 & $3.730 \times 10^{-5}$ & 5.832 & $1.702 \times 10^{-3}$ & 1.825 & $1.665 \times 10^{-4}$ & 1.573 & $1.238 \times 10^{-4}$ \\
\bottomrule
\end{tabularx}%
}
\end{adjustwidth}
\end{table}

\vspace{-5pt}
All selected fits in Table~\ref{tab:inclination-results} pass the sampled no-inversion and frequency-resolution checks described in Appendix~\ref{app:no-inversion-search}. A small residual demonstrates that an allowed no-inversion waveform closely reproduces the injection. A larger best-found residual is an upper bound on the minimum attainable within the allowed template family, because further search could increase the match.

\begin{figure}[H]
\begin{adjustwidth}{-\extralength}{0cm}
\centering
\includegraphics[width=0.96\linewidth]{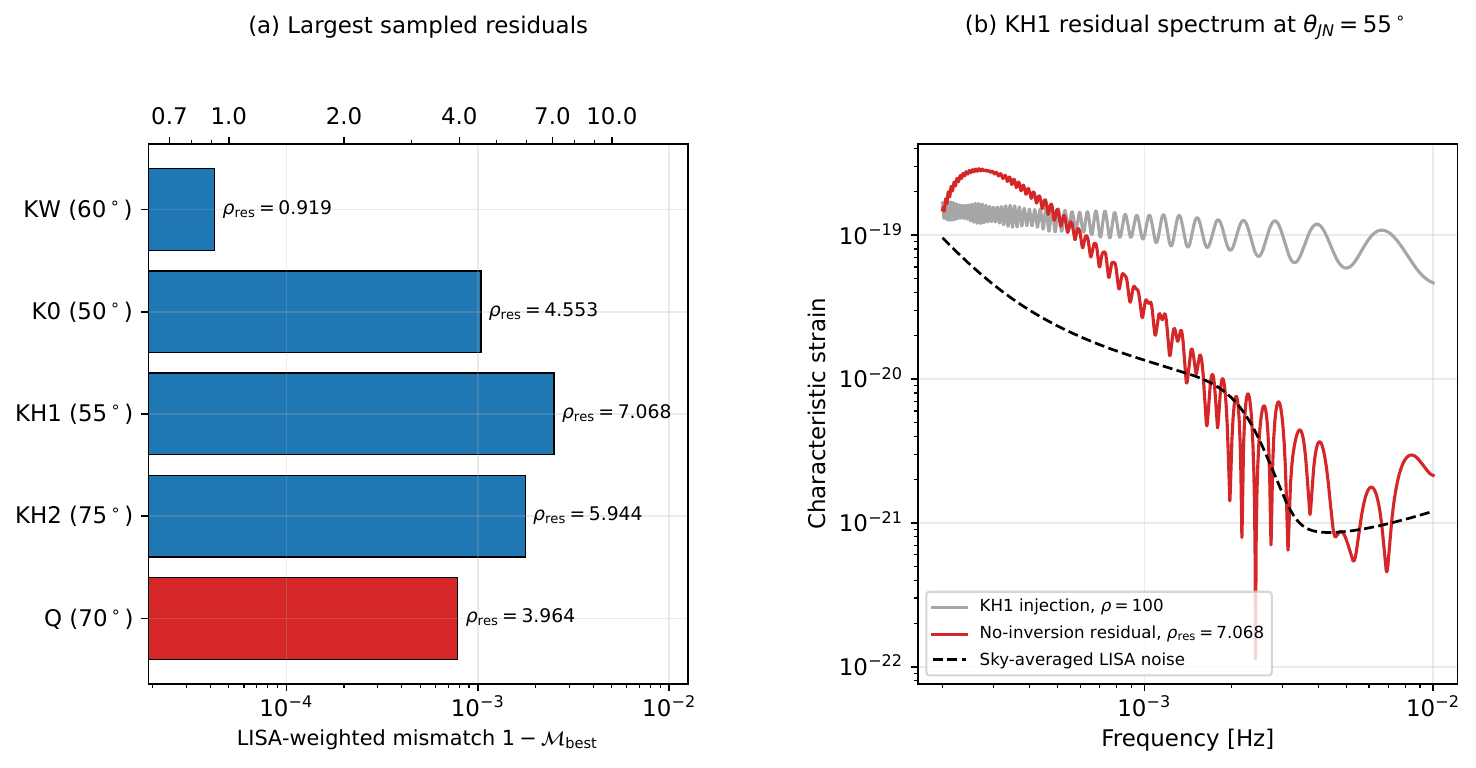}
\end{adjustwidth}
\caption{No-inversion recovery at the case-specific sampled residual maxima.%
Panel (\textbf{a}) shows the LISA-weighted mismatch of the selected best-found fit at
the inclination at which each case reaches its largest value in Table~\ref{tab:inclination-results}.
The upper scale gives the corresponding residual SNR at $\rho_\star=100$.
Panel (\textbf{b}) shows the characteristic strain of the KH1 injection at
$\theta_{JN}=55^\circ$ and its residual after subtraction of the corresponding
selected no-inversion waveform. Its residual SNR is $7.068$.
The dashed curve is the sky-averaged LISA noise amplitude, including the
one-year Galactic confusion foreground.}
\label{fig:observational-recovery}
\end{figure}

For the spectral comparison, let
\begin{equation}
 C=\frac{\rho_\star}{\rho[h_{\rm inv}]}
\end{equation}
rescale the injection to the reference SNR. After applying the fitted
amplitude $a$, phase shift $\Delta\phi$, and time shift $\Delta t$
to the best no-inversion waveform, the frequency-domain residual is
\begin{equation}
 \widetilde h_{\rm res}(f)
 =C\left[
 \widetilde h_{\rm inv}(f)
 -a e^{i\Delta\phi}e^{-2\pi i f\Delta t}
 \widetilde h_{\rm NI}(f)
 \right].
\end{equation}
The characteristic strains plotted in
Figure~\ref{fig:observational-recovery} are
\begin{equation}
 h_{c,\rm inv}(f)=2f\left|C\widetilde h_{\rm inv}(f)\right|,
 \qquad
 h_{c,\rm res}(f)=2f\left|\widetilde h_{\rm res}(f)\right|,
\end{equation}
and the plotted noise amplitude is
\begin{equation}
 h_n(f)=\sqrt{fS_n(f)}.
\end{equation}

\noindent
These curves show the frequency distribution of the signal and residual,
whereas the residual SNR is obtained by integrating over the full
frequency interval. Figure~\ref{fig:observational-recovery}b shows where the KH1 residual
lies in frequency.  The residual characteristic strain contains several
minima because the aligned complex waveforms cancel more closely at those
frequencies.  Its SNR is nevertheless determined by the integral over the
whole band.  The curve should therefore not be interpreted by comparing the
residual and noise at only one frequency.

\vspace{-3pt}

\vspace{-12pt}
\subsection{Quadrupole-Induced Inversions}
\label{subsec:quadrupole-inversions}

\begin{figure}[H]
\begin{adjustwidth}{-\extralength}{0cm}
\centering %
\includegraphics[scale=0.65]{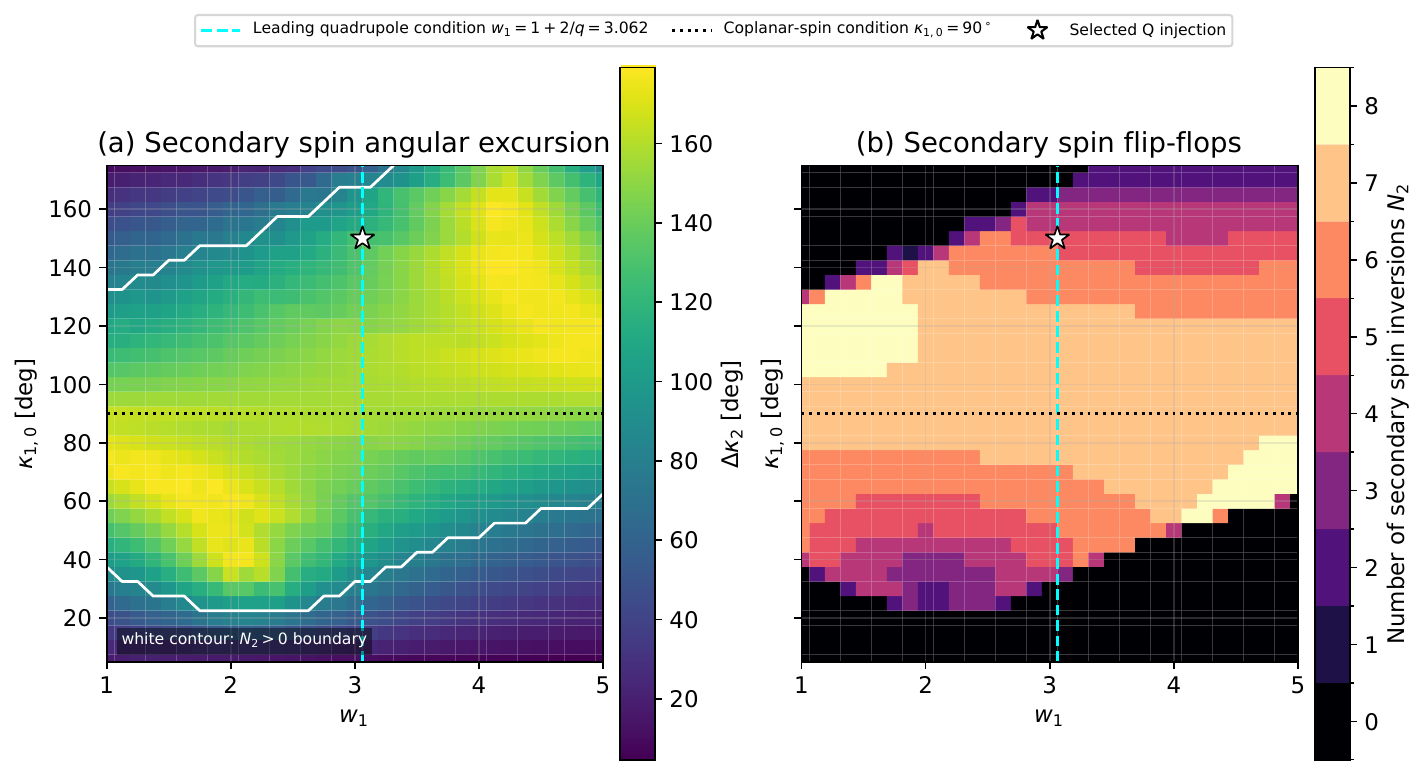}\end{adjustwidth}
\caption{Quadrupole-induced secondary-spin inversions. %
The scan varies $w_1$ and $\kappa_{1,0}$ at
\mbox{$M=2\times10^5M_\odot$,} $q=0.97$, $\chi_1=0.70$, $\chi_2=0.05$,
$\kappa_{2,0}=18^\circ$, $\Delta\zeta_0=45^\circ$, and $w_2=3$, over
$0.2\,{\rm mHz}\leq f\leq10\,{\rm mHz}$.  Panel (\textbf{a}) shows
$\Delta\kappa_2=\kappa_{2,\max}-\kappa_{2,\min}$; the white contour marks the
boundary $N_2>0$.  Panel (\textbf{b}) shows the number $N_2$ of crossings of
$\cos\kappa_2=0$.  The black dotted line marks $\kappa_{1,0}=90^\circ$, the
cyan dashed line marks the leading estimate $w_1\simeq1+2/q=3.062$, and the
star denotes the Q injection.}
\label{fig:quadrupole-scan}
\end{figure}

The Q configuration tests the quadrupole-induced case identified in
Ref.~\cite{KeresztesTapaiGergely2021}.  In the weak-secondary-spin limit and
near equal masses, its leading-order estimate is
\mbox{$w_1\simeq1+2/q$.}  For $q=0.97$, this gives $w_1\simeq3.0619$. Figure~\ref{fig:quadrupole-scan} shows the mapping of $1\leq w_1\leq5$ and $5^\circ\leq\kappa_{1,0}\leq175^\circ$, with the other
parameters fixed to those of Q.  The broad band around
$\kappa_{1,0}=90^\circ$ is the coplanar-spin flip-flop branch.  Away from this
band, a second region of large excursions and repeated crossings appears near
$w_1\simeq3$.  The selected Q injection lies in this second region; its
dominant spin remains far from the orbital plane, while its secondary crosses
the plane five times.  Thus, its motion is not the continuation of the
coplanar-spin branch at $\kappa_{1,0}=90^\circ$.

\begin{figure}[H]
\includegraphics[width=0.7\textwidth]{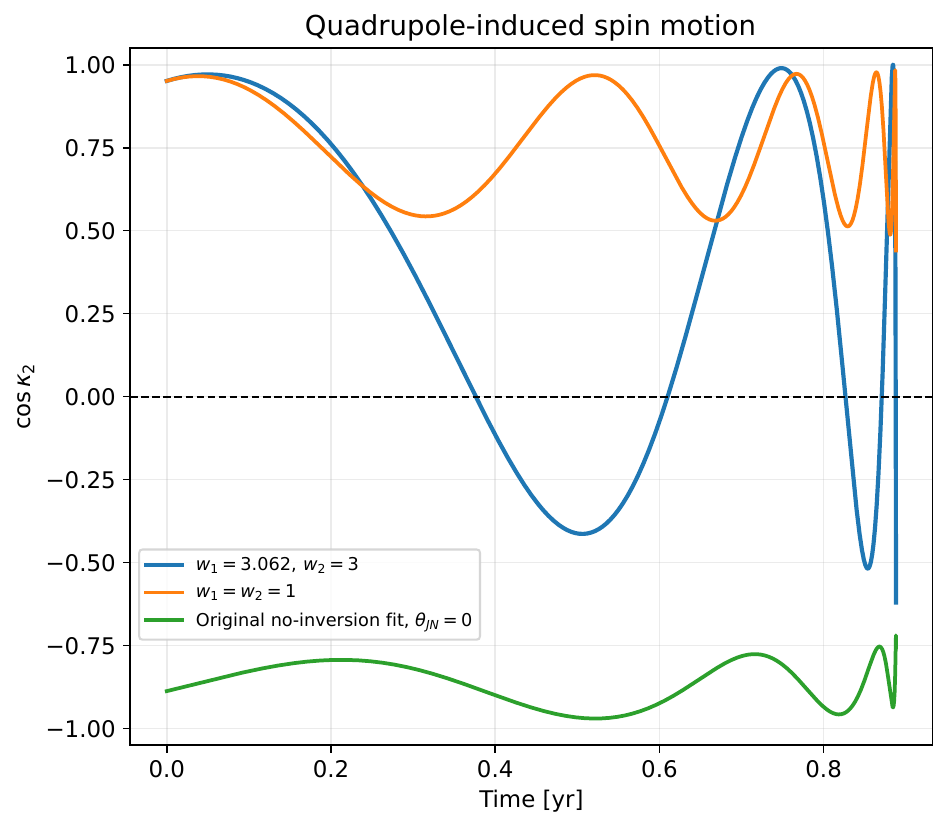}
\caption{Quadrupole-induced spin inversion. %
The blue curve shows the Q injection with \mbox{$w_1=3.0619$} and $w_2=3$.  The orange
curve has the same masses, spin magnitudes, and initial angles but Kerr
coefficients $w_1=w_2=1$. The green curve is the aligned-view
($\theta_{JN}=0$) no-inversion fit in Equation~\eqref{eq:q-best-fit}, with the Q
quadrupole coefficients retained. The plotted quantities describe intrinsic
spin evolution.}
\label{fig:quadrupole-comparison}
\end{figure}

Changing the quadrupole coefficients can change the dynamics significantly.  As
shown in Figure~\ref{fig:quadrupole-comparison}, the Q secondary repeatedly
crosses the orbital plane, whereas the otherwise identical system with Kerr
coefficients $w_1=w_2=1$ remains on the aligned side.  This establishes that
the Q trajectory is generated by the non-Kerr quadrupole terms.  It does not,
however, establish that the crossing can be identified from the restricted
waveform. The aligned-view no-inversion fit to Q lies in the $(-,-)$ sector and has
\begingroup\makeatletter\def\f@size{9}\check@mathfonts
\def\maketag@@@#1{\hbox{\m@th\normalsize \normalfont#1}}%
\begin{equation}
 (q,\chi_1,\chi_2)=(0.96459,\,0.25528,\,0.56456),\qquad
 (\kappa_{1,0},\kappa_{2,0},\Delta\zeta_0)
 =(112.90^\circ,\,152.49^\circ,\,35.46^\circ).
 \label{eq:q-best-fit}
\end{equation}\endgroup
\vspace{-25pt}
\vspace{-12pt}

The %
quadrupole coefficients remain fixed at $w_1=3.0619$ and $w_2=3$.  Both
spins remain on the anti-aligned side of the orbit, yet the waveform reaches
$\mathcal M_{\rm best}=0.9999897071$.  Its residual is only
$\rho_{\rm res}=0.45$ at $\rho_\star=100$.  The parameter changes in
Equation~\eqref{eq:q-best-fit} therefore reproduce the carrier phase and leading
inclination-amplitude evolution without reproducing the spin inversion itself. As a control, we repeated the Q spin evolution with $w_2=1$ while
retaining $w_1=3.0619$. The secondary-spin excursion changes from
$128^\circ$ to $126^\circ$, while the number of orbital-plane
crossings remains $N_2=5$. The quadrupole-induced behavior is therefore
controlled primarily by $w_1$ for this weak-secondary-spin~configuration.

\begin{figure}[H]
\includegraphics[width=0.85\textwidth]{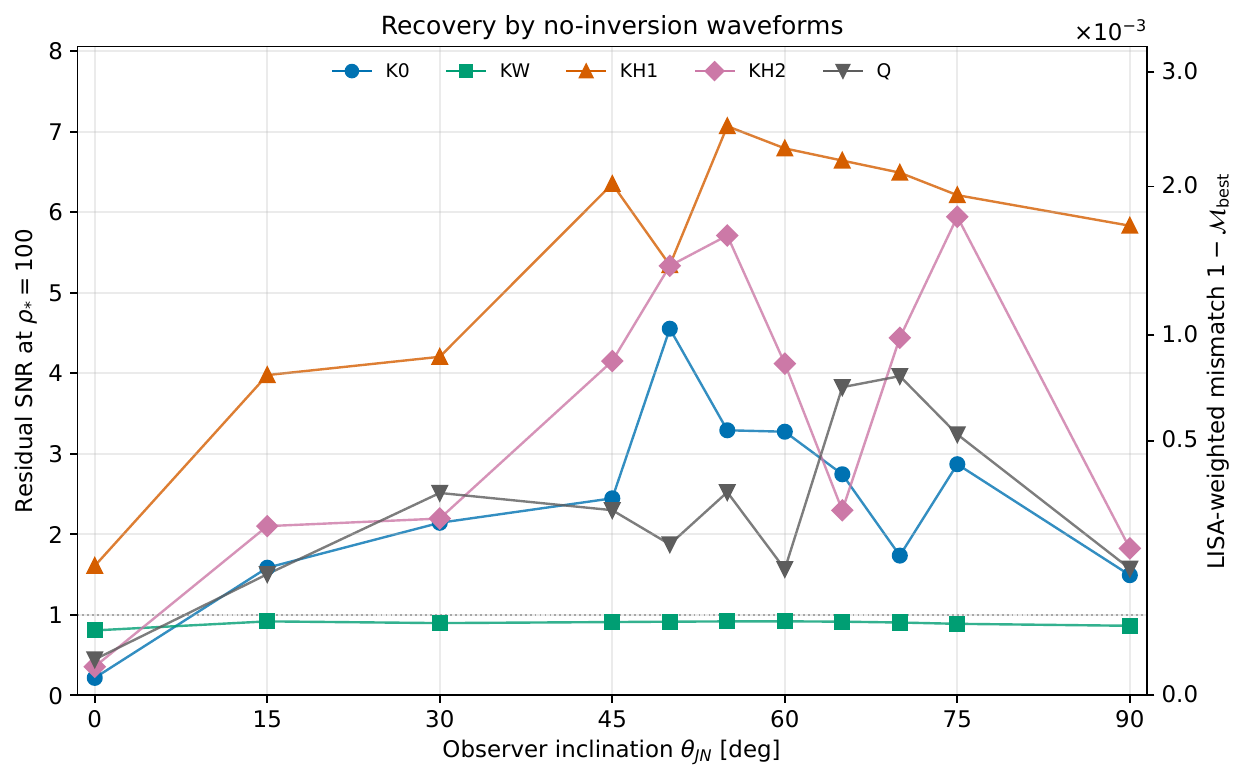}
\caption{Dependence of the selected no-inversion residual on observer inclination. %
The curves show the residual SNR
for K0, KW, KH1, KH2, and Q at eleven sampled values of $\theta_{JN}$.
Every injection is separately normalized to signal SNR $100$ at each direction,
and its no-inversion template is searched at that same fixed inclination.
The zero-degree points are the independently searched inclination-study fits,
rather than the fit set in Table~\ref{tab:no-inversion-results}.
The right-hand scale gives the corresponding LISA-weighted mismatch.
Markers distinguish the five cases, and connecting lines guide the eye between
sampled directions; they do not represent statistical uncertainty.
The search status is described in Appendix~\ref{app:no-inversion-search}.
All plotted residuals are best-found upper bounds on the minimum residual
within the allowed template family.}
\label{fig:inclination-dependence}
\end{figure}

The $\theta_{JN}=0^\circ$ results show that large and repeated spin inversions
can be hidden by parameter degeneracies in a waveform that contains only the
dominant harmonic, the accumulated 2PN carrier phase, and the leading
inclination-amplitude factor. The residual scales linearly with the chosen
reference SNR, but increasing $\rho_\star$ does not remove the limitations of
the waveform model or replace parameter estimation. A complete precessing
waveform adds inertial-frame mode rotation, precession sidebands, higher
harmonics, two polarization responses, and the time-dependent LISA response.
Those observables may distinguish trajectories that are degenerate here;
testing that possibility requires the complete waveform and a full recovery
analysis. Figure~\ref{fig:inclination-dependence} shows the selected residuals across the sampled inclinations.

\vspace{-3pt}

\section{Concluding Remarks}

In this work, we study whether the spin inversions generated by the secular dynamics
leave a waveform difference that cannot be reproduced by a physically evolving
binary without an orbital-plane crossing. For $\theta_{JN}=0^\circ$, with the line of sight along $\bm J(t_0)$,
the five near-equal-mass injections are closely reproduced by
no-inversion waveforms at the reference signal-to-noise ratio
$\rho_\star=100$. After
searching all possible no-inversion sectors and varying the detector-frame total
mass, mass ratio, spin magnitudes, and initial spin angles, the residual SNRs
are $0.75$ for K0, $0.81$ for KW, $1.64$ for KH1, $0.73$ for KH2, and
$0.45$ for Q.  The corresponding matches all exceed $0.999865$.  The residual SNR values
are the smallest numerically found upper bounds.

The inclination extension gives a more qualified conclusion. At eleven sampled
directions between $0^\circ$ and $90^\circ$, the KW residual remains below
unity when each injection is normalized to $\rho_\star=100$. For K0, KH1,
KH2, and Q, the largest selected best-found residuals among those directions
are $4.55$, $7.07$, $5.94$, and $3.96$, respectively. Their maxima occur at
different sampled inclinations. The selected fits are stable under increased
frequency resolution, while search convergence is not yet fully established.
Their larger residuals are upper bounds on the minimum achievable in the
allowed template family. This comparison fixes the template inclination to
that of the injection and restores only the restricted inclination-amplitude
modulation. At $\rho_\star=100$, residual SNRs of $4$--$7$ lie well above the
unit-residual scale for waveform indistinguishability. They indicate that, at
some nonzero inclinations, spin-inverting signals may contain measurable
structure not reproduced by the selected no-inversion waveforms.

The $\theta_{JN}=0^\circ$ results show why the angular size of an inversion
is not a measure of its observability.  In KW, the weak secondary spin spans
$133^\circ$ and crosses the orbital plane nine times, but the best
no-inversion residual is only $\rho_{\rm res}=0.81$.  In KH2, both spins
move through approximately $139^\circ$, yet the residual decreases to $0.73$.
KH1 contains fourteen crossings of each spin and gives the largest residual in
the sample, but even there $\rho_{\rm res}=1.64$.  A large spin magnitude,
a large angular excursion, and many orbital-plane crossings are therefore not
sufficient conditions for a distinguishable restricted waveform.

The quadrupole calculation establishes a separate dynamical result. Scanning
$w_1$ and $\kappa_{1,0}$ reveals both the coplanar-spin branch near
$\kappa_{1,0}=90^\circ$ and a second region of secondary-spin inversions near
the leading quadrupole condition $w_1\simeq1+2/q$.  The selected Q injection,
with $q=0.97$, $w_1=3.0619$, and $w_2=3$, lies in the latter region.  Its
dominant spin remains far from the orbital plane, while its weak secondary
spans $128^\circ$ and crosses the plane five times.  Replacing both
quadrupole coefficients by their Kerr values removes these crossings for the
otherwise identical source.  The inversion is therefore produced by the
non-Kerr quadrupole terms rather than by the coplanar-spin mechanism. At $\theta_{JN}=0^\circ$, this difference is nevertheless almost completely absorbed by
the no-inversion fit.  Keeping $w_1=3.0619$ and $w_2=3$, a source in the
$(-,-)$ sector reaches $\mathcal M_{\rm best}=0.9999897071$ with Q and leaves
$\rho_{\rm res}=0.45$ at $\rho_\star=100$.  Its two spins remain on the
anti-aligned side of the orbit throughout the calculation.  Thus, the
restricted waveform can reproduce the carrier phase and leading
inclination-amplitude change without reproducing the quadrupole-induced
inversion.  The value $w_1\simeq3$ is used here only to test the quadrupole-induced dynamics at a common mass and LISA frequency scale, not as a specific astrophysical model of these~values.

The observational conclusion is limited by the waveform model.  The
calculation assumes a circular inspiral, where precession enters the carrier phase
through the evolving $\beta$ and $\sigma$ coefficients and enters the leading
amplitude through the inclination factor.  The
noise weighting is sky averaged and includes a one-year Galactic confusion
foreground. A complete analysis should combine the secular inversion dynamics with a
precessing inspiral--merger--ringdown waveform and retain the separate
polarizations, inertial-frame mode rotation, precession sidebands, higher
harmonics, and moving LISA response.  It should then compare each
spin-inverting signal with the best ordinary-precession explanation over the
full source parameter space, using global optimization or Bayesian inference.
Eccentricity and the unequal-mass supermassive-black-hole regime should be
examined separately, and the comparable-mass cases should be connected to
numerical-relativity and effective-one-body descriptions
~\cite{VarmaEtAl2021,GerosaEtAl2023}.  These additional observables may break the
degeneracies found by the restricted carrier waveform.  The injections and
no-inversion searches presented here provide concrete cases for such a test.
Their spin transitions are dynamically well defined, but identifying them
in LISA data remains an open waveform and parameter-recovery problem.

In a prospective LISA analysis, the full-band comparison provides the first-stage test as the occurrence and timing of a spin inversion are not known in advance. If a candidate inversion is identified, a local comparison provides a targeted second-stage test. For the five $\theta_{JN}=0^\circ$ fits, restricting the analysis either to the full inversion-containing interval or to individual inversion-containing windows does not increase the absolute residual SNR. The normalized mismatch can nevertheless increase within an individual window, showing that localization can enhance the fractional waveform contrast even when the reduced signal content of the short interval prevents a larger absolute residual. Full local parameter recovery with a complete precessing LISA waveform is left to future studies.

\vspace{6pt}
\authorcontributions{Conceptualization, L.Á.G.; methodology, K.K. and L.Á.G.; software, K.K. and L.Á.G.; validation, K.K.; formal analysis, K.K. and L.Á.G.; investigation, K.K. and L.Á.G.; data curation, K.K.; writing---draft preparation, L.Á.G.; writing---review and editing, K.K. and L.Á.G.; visualization, K.K.; supervision, L.Á.G.; project administration, L.Á.G.; funding acquisition, L.Á.G. All authors have read and agreed to the published version of the manuscript.}

\funding{This research was funded by the HUN-REN Wigner Research Centre for Physics, grant number RMI 39000-03, “Gravitational waves and their sources”.}

\dataavailability{The source code, Jupyter notebook, numerical data, and scripts for the
aligned-view calculation are openly available in Zenodo, version
0.1.0, at \url{https://doi.org/10.5281/zenodo.22025538}. The observer-inclination
extension, including the notebooks, selected fits, saved search candidates,
refinement records, and numerical verification files, is available in version
0.2.0 at \url{https://doi.org/10.5281/zenodo.23309463}.}

\acknowledgments{We thank the referees for constructive suggestions. We acknowledge the computational resources provided by the Wigner Scientific Computing Laboratory (WSCLAB) at the HUN-REN Wigner Research Centre for Physics.}
\conflictsofinterest{The authors declare no conflicts of interest. %
 }

\appendixtitles{yes}
\appendixstart
\appendix
\section{No-Inversion Search and Numerical Verification}
\label{app:no-inversion-search}

\subsection{Aligned-View Searches}

This subsection describes how the aligned-view no-inversion waveforms
were selected and how their matches were checked numerically. For each injection, we searched the
four sectors $(+,+)$, $(+,-)$, $(-,+)$, and $(-,-)$. A plus sign means that
the corresponding spin projection remains positive throughout the frequency
interval, while a minus sign means that it remains negative. For example, the
sector $(-,+)$ requires
$\cos\kappa_1<0$ and $\cos\kappa_2>0$ throughout the evolution. Every
candidate waveform was evolved with the same secular equations and frequency
evolution as its corresponding injection.

For sector signs $(s_1,s_2)$, where each $s_i$ is $+1$ or $-1$, the minimum
signed projections $\mu_i$ are defined in
Equation~\eqref{eq:no-inversion-margin}. A candidate was rejected if
\begin{equation}
\mu_i<10^{-4}
\label{eq:no-inversion-rejection}
\end{equation}
for either spin. This condition does more than exclude a detected crossing:
it also requires the accepted waveform to remain a small but finite distance
from the orbital plane on the numerical frequency grid.

For every accepted candidate, the correlation was maximized over a time shift
\mbox{$\Delta t\in[-10^4,10^4]\,{\rm s}$.} The absolute value of the complex
correlation performs the maximization over a constant phase shift. Once the
best time and phase shifts have been found, the constant non-negative
amplitude factor that gives the smallest residual is

\begin{equation}
a_{\rm best}
=\frac{|z|}{(h_{\rm NI}|h_{\rm NI})},
\label{eq:best-amplitude-factor}
\end{equation}
where $h_{\rm NI}$ is the no-inversion waveform. After rescaling the injection
to the reference SNR $\rho_\star=100$, the SNR remaining in the residual is
\begin{equation}
\rho_{\rm res}
=\rho_\star\sqrt{1-\mathcal M_{\rm best}^{2}}.
\label{eq:best-residual-snr-app}
\end{equation}
When $\mathcal M_{\rm best}$ is close to unity, this becomes
$\rho_\star\sqrt{2(1-\mathcal M_{\rm best})}$ to leading order in \mbox{the
mismatch.}

The spherical angles become inconvenient when a spin approaches either
orbital pole because its azimuth is then undefined. We therefore do not
integrate $(\kappa_1,\kappa_2,\Delta\zeta)$ directly. Instead, the numerical
calculation uses
\begin{equation}
c_i=\cos\kappa_i,\qquad
c_\gamma=\cos\gamma,\qquad
p=\sin\kappa_1\sin\kappa_2\sin\Delta\zeta.
\end{equation}
Defining
\begin{equation}
u=c_\gamma-c_1c_2
=\sin\kappa_1\sin\kappa_2\cos\Delta\zeta,
\end{equation}
the relative azimuth can be reconstructed after the integration as
\begin{equation}
\Delta\zeta=\operatorname{atan2}(p,u).
\end{equation}
This choice avoids the divisions by $\sin\kappa_i$ that make the direct
equation for $d\Delta\zeta/d\tau$ singular in spherical coordinates when
a spin approaches the orbital axis. As a check that these
variables continue to describe a consistent spin configuration, we monitored
the algebraic relation
\begin{equation}
p^2+u^2=(1-c_1^2)(1-c_2^2).
\label{eq:numerical-angular-constraint}
\end{equation}
The maximum absolute violation of this relation among the five injections was
$1.6\times10^{-9}$.

The search varied the detector-frame total mass $M$, the mass ratio $q$, the
spin magnitudes $\chi_1$ and $\chi_2$, and the initial angles
$\kappa_{1,0}$, $\kappa_{2,0}$, and $\Delta\zeta_0$. The quadrupole
coefficients were fixed to the values of the corresponding injection:
$w_1=w_2=1$ for K0, KW, KH1, and KH2, and
$(w_1,w_2)=(3.0619,3)$ for Q. The total-mass interval corresponds to allowing
the no-inversion mass to differ from the injection mass by approximately
$2\%$. Table~\ref{tab:no-inversion-search-ranges} gives the complete search
intervals.
\vspace{-3pt}
\begin{table}[H]
\caption{Parameter ranges used in the aligned-view no-inversion
searches. The inclination extension uses the same intrinsic ranges and adds
the initial orbital azimuth at nonzero viewing inclination.}
\label{tab:no-inversion-search-ranges}
\small
\begin{tabularx}{\textwidth}{lll}
\toprule
\textbf{Parameter} & \textbf{Search Interval} & \textbf{Treatment} \\
\midrule
$\ln(M_{\rm NI}/M_{\rm inj})$
& $[-0.02,\,0.02]$
& varied in stage 2 \\
$q$
& $[0.50,\,1.00]$
& varied \\
$\chi_1,\chi_2$
& $[0,\,0.999]$
& varied \\
$\kappa_{i,0}$ for $s_i=+1$
& $[0.1^\circ,\,89.9^\circ]$
& varied \\
$\kappa_{i,0}$ for $s_i=-1$
& $[90.1^\circ,\,179.9^\circ]$
& varied \\
$\Delta\zeta_0$
& $[0^\circ,\,360^\circ]$
& varied \\
$w_1,w_2$
& injection values
& fixed \\
$\Delta t$
& $[-10^4,\,10^4]\,{\rm s}$
& maximized numerically \\
$\Delta\phi$
& $[0,\,2\pi)$
& maximized through the complex correlation \\
Overall amplitude $a$
& $a\geq0$
& fitted analytically \\
\bottomrule
\end{tabularx}
\end{table}
\vspace{-5pt}
The optimization was carried out in two stages. Stage 1 varied $q$,
$\chi_1$, $\chi_2$, and the three initial angles, while keeping the total mass
equal to the injection mass. Stage 2 also allowed the total mass to vary and
started from the best candidate found in Stage 1. Before each
differential-evolution search, we evaluated a fixed set of physically allowed
starting configurations and retained the best one as an initial candidate.
The second differential-evolution stage was followed by a local Powell
refinement. During the global search, the correlation was first evaluated on a grid of
time shifts covering $[-10^4,10^4] {\rm s}$ with a spacing of $50\,{\rm s}$.
A bounded one-dimensional maximization then refined the best grid point. For
the final verification, we used an $801$-point grid over the same interval,
corresponding to a spacing of $25\,{\rm s}$, followed by another continuous
refinement.

Two search depths were used. The moderate search evaluated each waveform on
$512$ frequency samples. Its two differential-evolution stages were limited
to $8$ and $12$~generations, the SciPy \texttt{popsize%
} parameter was $4$,
and the random seed was $20260818$. The intensive search used $1024$
frequency samples, limits of $40$ and $60$ generations,
\texttt{popsize}$=8$, and two independent random seeds, $20260818$ %
 and
$20260917$. For every final candidate, both the injection and the fitted
waveform were then reintegrated on $8192$ frequency samples with a relative
integration tolerance of $10^{-10}$. The aligned-view matches and residual SNRs in
Table~\ref{tab:no-inversion-results} come from this high-resolution verification.

Table~\ref{tab:best-no-inversion-parameters} lists the physically evolving
no-inversion waveforms that produced the largest matches found. In every case,
the fitted total mass returned to the injection value
$M=2\times10^5M_\odot$. The corresponding sectors, matches, no-inversion
margins, and residual SNRs are given in
Table~\ref{tab:no-inversion-results}.
Digits retained in the optimization tables are provided for numerical reproducibility and should not be interpreted as observational \mbox{parameter precision.}

\begin{table}[H]
\caption{Parameters of the aligned-view no-inversion waveforms in
Table~\ref{tab:no-inversion-results}. All fitted total masses are
$M=2\times10^5M_\odot$. These parameter vectors are retained for the window comparisons and Q~illustration.}
\label{tab:best-no-inversion-parameters}
\begin{tabularx}{\textwidth}{LCCC ccc CC}
\toprule
\textbf{Case} & \boldmath{$q$} & \boldmath{$\chi_1$} & \boldmath{$\chi_2$}
& \boldmath{$\kappa_{1,0}$} \textbf{(deg)} & \boldmath{$\kappa_{2,0}$} \textbf{(deg)} & \boldmath{$\Delta\zeta_0$} \textbf{(deg)}
& \boldmath{$w_1$} & \boldmath{$w_2$} \\
\midrule
K0
& 0.9700 & 0.5000 & 0.5000
& 75.00 & 45.00 & 180.00 & 1 & 1 \\
KW
& 0.9700 & 0.0500 & 0.9500
& 75.00 & 88.00 & 0.00 & 1 & 1 \\
KH1
& 0.9486 & 0.1039 & 0.6480
& 169.22 & 71.81 & 170.78 & 1 & 1 \\
KH2
& 0.9778 & 0.0764 & 0.0630
& 44.65 & 24.48 & 51.87 & 1 & 1 \\
Q
& 0.9646 & 0.2553 & 0.5646
& 112.90 & 152.49 & 35.46 & 3.0619 & 3 \\
\bottomrule
\end{tabularx}
\end{table}
\vspace{-5pt}
All aligned-view final fits return to the injected total mass. The corresponding chirp masses differ from the injections by at most $0.016\%$. At the initial frequency, the effective projected-spin combination $\chi_{\rm eff,0}=[\chi_1\cos\kappa_{1,0}+q\chi_2\cos\kappa_{2,0}]/(1+q)$ changes by at most about $0.036$ in absolute value across the five cases. The individual spin magnitudes and orientations can nevertheless shift much more strongly. Within the restricted model, the degeneracy is therefore absorbed mainly by changes in the detailed spin configuration rather than by a large chirp-mass bias.

Table~\ref{tab:search-depth} shows why the more extensive search was needed
for KH1, KH2, and Q. In each of these cases, the intensive search found a
no-inversion waveform with a larger match. A better match leaves a smaller
waveform residual, so the corresponding residual SNR also decreased. The Q
comparison keeps $w_1=3.0619$ and $w_2=3$ fixed in both searches. Earlier
exploratory calculations in which $w_1$ was allowed to vary are not included
in the reported~results.

K0 and KW were not repeated with the intensive settings because the moderate
searches had already found valid no-inversion waveforms with
$\rho_{\rm res}<1$ at $\rho_\star=100$. The existence of these
high-matching no-inversion waveforms is already sufficient to demonstrate a
strong degeneracy within the restricted model. A more extensive search could
leave the best match unchanged or increase it; it could not establish a
larger minimum residual. By contrast, the moderate KH1, KH2, and Q residuals
were large enough that deeper searches were needed before they could be
interpreted. For every final candidate, the high-resolution calculation gave $\mu_i>0$
and zero detected sign changes of both spin projections. The smallest final
margin was $\mu_i=2.42\times10^{-2}$, which is more than two~orders of
magnitude above the rejection threshold. The accepted no-inversion waveforms
therefore remain well-separated from the orbital plane on the adopted
frequency grid. To count crossings in the spin-inverting injections, we first removed samples
inside the narrow numerical band
\begin{equation}
|\cos\kappa_i|\leq10^{-6}
\end{equation}
and then counted sign changes between the remaining consecutive samples. We
did not use a continuous root-finding procedure to locate each crossing.
Consequently, the reported counts for trajectories that start at, or pass
very close to, $\cos\kappa_i=0$ depend on this sampling rule. This applies in
particular to the primary-spin trajectories in K0 and KW. Their crossing
counts should therefore not be interpreted as numbers of complete
flip-flop cycles.

\vspace{-3pt}
\begin{table}[H]
\centering
\caption{Dependence of the aligned-view no-inversion recovery on numerical search
depth. The residual SNRs are evaluated at $\rho_\star=100$. A smaller
intensive-search residual means that the moderate search did not locate the
highest-match region subsequently found. The intensive values are the largest
matches obtained numerically and are not proofs of the exact global maxima.}
\label{tab:search-depth}
\begin{tabularx}{\textwidth}{L c C c C c}
\toprule
\textbf{Case}
& \boldmath{$\mathcal M_{\rm mod}$}
& \boldmath{$\rho_{\rm res,mod}$}
& \boldmath{$\mathcal M_{\rm int}$}
& \boldmath{$\rho_{\rm res,int}$}
& \textbf{Residual Reduction} \\
\midrule
KH1
& 0.9982692956
& 5.881
& 0.9998652957
& 1.641
& $72\,\%$ \\
KH2
& 0.9993163960
& 3.697
& 0.9999731354
& 0.733
& $80\,\%$ \\
Q
& 0.9991862150
& 4.033
& 0.9999897071
& 0.454
& $89\,\%$ \\
\bottomrule
\end{tabularx}
\end{table}

\vspace{-5pt}
Finally, we checked whether increasing the number of frequency samples changes
the aligned-view results. The final fixed-parameter comparisons were repeated
with $4096$, $8192$, and 16,384 samples. Between the $8192$- and
16,384-point calculations, the largest absolute change in
$\mathcal M_{\rm best}$ was $2.9\times10^{-11}$, and the largest change in
$\rho_{\rm res}$ was $1.8\times10^{-7}$ at $\rho_\star=100$. The
relative change in the mismatch was at most $2.4\times10^{-7}$. Every fitted
waveform still had zero spin-projection crossings, and the smallest
no-inversion margin at 16,384 samples remained
$2.42\times10^{-2}$. These changes are far too small to affect the physical
conclusions. The reported matches, residual SNRs, and no-inversion
classifications are therefore stable under this increase in frequency
resolution.

\subsection{Inclination-Dependent Searches and Verification}
\label{app:inclination-search}

The inclination extension retains the intrinsic search ranges in
Table~\ref{tab:no-inversion-search-ranges} and the sampled signed-margin
requirement in Equation~\eqref{eq:no-inversion-rejection}. For every case and
inclination, all four no-inversion sectors were searched independently.
The injected initial azimuth is $\lambda_{L,0}=0$, and the template azimuth
is varied over $[0,2\pi)$ at nonzero $\theta_{JN}$. Both are fixed to zero
at $\theta_{JN}=0$, where this angle does not affect the amplitude.
The template inclination equals the injection inclination. The nodal phase is integrated by cumulative
trapezoidal quadrature along the evolved $\tau(f)$.

The standard searches used $1024$ frequency samples, differential-evolution
limits of $40$ and $60$ generations in the two stages, population multiplier
$8$, and seeds $20260818$ and $20260917$. For nonzero-inclination KH1, the
limits were $80$ and $120$ generations, the population multiplier was $12$,
and the seeds were $20260818$, $20260917$, $20261007$, and $20261008$.
The aligned-view parameter vectors were used for regression
comparisons, not as starting points for these independent searches.
At zero inclination, two additional KH1 searches with seeds $20261007$ and
$20261008$ covered the $(-,+)$ sector using the $80/120$-generation and
population-multiplier-$12$ settings. Ten subsequent local-refinement attempts
started from saved candidates for KH1 at $60^\circ$ and $90^\circ$ and Q at
$60^\circ$.

Final comparisons reintegrated both injection and template with relative
tolerance $10^{-10}$ and maximized the time and phase alignment. At nonzero
inclination, the template initial node was profiled on $36$- and $144$-point
grids with continuous local refinement where used, while preserving the saved
time and node alignments and earlier verified alignments whenever they gave a
larger match. The best admissible saved candidate was selected across sectors
and seeds after this alignment verification. Selection and alignment
verification do not constitute additional intrinsic searches.

The final selection contains $55$ fits and $165$ evaluations at $4096$,
$8192$, and 16,384 frequency samples. All have zero sampled template-spin
crossings and satisfy the signed-margin constraint. The smallest signed
margin across these evaluations is $8.9563\times10^{-4}$. The largest match spread over the three resolutions
is $3.8167\times10^{-8}$. Between $8192$ and 16,384 samples, the largest
match change is $7.6318\times10^{-9}$ and the largest residual-SNR change is
$1.3063\times10^{-5}$ at $\rho_\star=100$. All five zero-degree regression
comparisons pass, but their selected parameter vectors are distinct from the
aligned-view fits. The resolution bounds above apply only to the fit set in
Table~\ref{tab:inclination-results}.

Frequency-resolution stability, sampled admissibility, regression against the
aligned-view calculation, and agreement between finite searches are separate
checks. A search warning is retained when independent-search or refinement
comparisons exceed a match spread of $10^{-5}$ or a residual-SNR spread of
$0.1$ at $\rho_\star=100$. The consolidated selection retains such warnings
for $42$ of the $55$ case--inclination groups. Agreement can also arise when different seeds select the same deterministic candidate.

The zero-degree, $60^\circ$, and $90^\circ$ searches and refinements used
WSCLAB resources; the eight additional inclinations were searched on a Mac Pro.
The $60^\circ$ and $90^\circ$ campaigns produced $96$ saved sector-search
records, and the eight additional inclinations produced $384$. In all $480$
records, the final differential-evolution stage reached its configured iteration
limit. The saved source, candidate records,
alignment checks, refinement history, and warning flags accompany the
inclination results.

\begin{adjustwidth}{-\extralength}{0cm}

\reftitle{References}

\PublishersNote{}
\end{adjustwidth}

\begin{thebibliography}{999}
\bibitem{GergelyBiermann2009} Gergely, L.\'A.; Biermann, P.L. The spin-flip phenomenon in supermassive black hole binary mergers. \emph{Astrophys. J.} \textbf{2009}, \emph{697}, 1621--1633. \url{https://doi.org/10.1088/0004-637X/697/2/1621}.

\bibitem{GergelyBiermannCaramete2010} Gergely, L.\'A.; Biermann, P.L.; Caramete, L.I. Supermassive black hole spin-flip during the inspiral. \emph{Class. Quantum Grav.} \textbf{2010}, \emph{27},~194009. \url{https://doi.org/10.1088/0264-9381/27/19/194009}.

\bibitem{LISA2023} Amaro-Seoane, P.; Andrews, J.; Arca Sedda, M.; Askar, A.; Baghi, Q.; Balasov, R.; Bartos, I.; Bavera, S.S.; Bellovary, J.; LISA Consortium; et al. Astrophysics with the Laser Interferometer Space Antenna. \emph{Liv. Rev. Relativ.} \textbf{2023}, \emph{26}, 2. \url{https://doi.org/10.1007/s41114-022-00041-y}.

\bibitem{LoustoHealy2015} Lousto, C.O.; Healy, J. Flip-flopping binary black holes. \emph{Phys. Rev. Lett.} \textbf{2015}, \emph{114}, 141101. %
  \url{https://doi.org/10.1103/PhysRevLett.114.141101}.
  
\bibitem{LoustoHealyNakano2016} Lousto, C.O.; Healy, J.; Nakano, H. Spin flips in generic black hole binaries. \emph{Phys. Rev. D} \textbf{2016}, \emph{93}, 044031. \url{https://doi.org/10.1103/PhysRevD.93.044031}.

\bibitem{KeresztesTapaiGergely2021} Keresztes, Z.; T\'apai, M.; Gergely, L.\'A. Spin and quadrupolar effects in the secular evolution of precessing compact binaries with black hole, neutron star, gravastar, or boson star components. \emph{Phys. Rev. D} \textbf{2021}, \emph{103}, 084024.  \url{https://doi.org/10.1103/PhysRevD.103.084024}.

\bibitem{GergelyBiermann2012} Gergely, L.\'A.; Biermann, P.L. The typical mass ratio and typical final spin in supermassive black hole mergers. \emph{arXiv} \textbf{2012}, arXiv:1208.5251. \url{https://doi.org/10.48550/arXiv.1208.5251}.

\bibitem{WangEtAl2025} Wang, B.Y.; Zhou, Y.; Chen, W.; Chen, N.; Di Matteo, T.; Croft, R.; Bird, S.; Ni, Y. Gravitational waves from massive black hole mergers in ASTRID: Predictions for LISA. \emph{Astrophys. J.} \textbf{2025}, \emph{993}, 199. \url{https://doi.org/10.3847/1538-4357/ae073e}.

\bibitem{LiaoEtAl2025} Liao, S.; Irodotou, D.; Maltz, M.G.A.; Lovell, C.C.; Jiang, Z.; Newman, S.L.; Vijayan, A.P.; Punyasheel, P.; Roper, W.J.; Seeyave, L.T.C.; et al. First light and reionization epoch simulations (FLARES)--XIX. Supermassive black hole mergers in the early Universe and their environmental dependence. \emph{Mon. Not. R. Astron. Soc.} \textbf{2025}, \emph{543}, 3055--3070. \url{https://doi.org/10.1093/mnras/staf1642}.

\bibitem{SesanaVolonteriHaardt2007} Sesana, A.; Volonteri, M.; Haardt, F. The imprint of massive black hole formation models on the LISA data stream. \emph{Mon. Not. R. Astron. Soc.} \textbf{2007}, \emph{377}, 1711--1716. \url{https://doi.org/10.1111/j.1365-2966.2007.11734.x}.

\bibitem{SayebEtAl2021} Sayeb, M.; Blecha, L.; Kelley, L.Z.; Gerosa, D.; Kesden, M.; Thomas, J. Massive black hole binary inspiral and spin evolution in a cosmological framework. \emph{Mon. Not. R. Astron. Soc.} \textbf{2021}, \emph{501}, 2531--2546. \url{https://doi.org/10.1093/mnras/staa3826}.

\bibitem{LoustoHealy2016} Lousto, C.O.; Healy, J. Unstable flip-flopping spinning binary black holes. \emph{Phys. Rev. D} \textbf{2016}, \emph{93}, 124074. \url{https://doi.org/10.1103/PhysRevD.93.124074}.

\bibitem{PislanEtAl2026} Pislan, F.-C.; Caramete, L.-I.; Caramete, A. Enhancing the scientific exploitation of future gravitational wave experiments through a multi-messenger approach. \emph{Class. Quantum Grav.} \textbf{2026}, \emph{43}, 085002. \url{https://doi.org/10.1088/1361-6382/ae59e3}.

\bibitem{KeresztesGergely2021} Keresztes, Z.; Gergely, L.\'A. Stability analysis of the spin evolution fixed points in inspiraling compact binaries with black hole, neutron star, gravastar, or boson star components. \emph{Phys. Rev. D} \textbf{2021}, \emph{103}, 084025. \url{https://doi.org/10.1103/PhysRevD.103.084025}.

\bibitem{MikocziVasuthGergely2005} Mik\'oczi, B.; Vas\'uth, M.; Gergely, L.\'A. Self-interaction spin effects in inspiralling compact binaries. \emph{Phys. Rev. D} \textbf{2005}, \emph{71}, 124043. \url{https://doi.org/10.1103/PhysRevD.71.124043}.

\bibitem{GergelyMikoczi2009} Gergely, L.\'A.; Mik\'oczi, B. Renormalized 2PN spin contributions to the accumulated orbital phase for LISA sources. \emph{Phys. Rev. D} \textbf{2009}, \emph{79}, 064023. \url{https://doi.org/10.1103/PhysRevD.79.064023}.

\bibitem{BarausseCardosoPani2014} Barausse, E.; Cardoso, V.; Pani, P. Can environmental effects spoil precision gravitational-wave astrophysics? \emph{Phys. Rev. D} \textbf{2014}, \emph{89}, 104059. \url{https://doi.org/10.1103/PhysRevD.89.104059}.

\bibitem{SteinleGerosa2023} Steinle, N.; Gerosa, D. The Bardeen--Petterson effect, disc breaking, and the spin orientations of supermassive black hole binaries. \emph{Mon. Not. R. Astron. Soc.} \textbf{2023}, \emph{519}, 5031--5042. \url{https://doi.org/10.1093/mnras/stac3821}.

\bibitem{GargEtAl2024} \textls[-15]{Garg, M.; Derdzinski, A.; Tiwari, S.; Gair, J.; Mayer, L. Measuring eccentricity and gas-induced perturbation from gravitational waves of LISA massive black hole binaries.} \emph{Mon. Not. R. Astron. Soc.} \textbf{2024}, \emph{532}, 4060--4074. \url{https://doi.org/10.1093/mnras/stae1764}.

\bibitem{Kidder1995} Kidder, L.E. Coalescing binary systems of compact objects to post-Newtonian $5/2$ order. V. Spin effects. \emph{Phys. Rev. D} \textbf{1995}, \emph{52}, 821--847. \url{https://doi.org/10.1103/PhysRevD.52.821}.

\bibitem{ApostolatosEtAl1994}
Apostolatos, T.A.; Cutler, C.; Sussman, G.J.; Thorne, K.S.
Spin-induced orbital precession and its modulation of the gravitational
waveforms from merging binaries.
\emph{Phys. Rev. D} \textbf{1994}, \emph{49}, 6274--6297.
\url{https://doi.org/10.1103/PhysRevD.49.6274}.

\bibitem{FairhurstEtAl2020} Fairhurst, S.; Green, R.; Hoy, C.; Hannam, M.; Muir, A. Two-harmonic approximation for gravitational waveforms from precessing binaries. \emph{Phys. Rev. D} \textbf{2020}, \emph{102}, 024055. \url{https://doi.org/10.1103/PhysRevD.102.024055}.

\bibitem{RobsonCornishLiu2019} Robson, T.; Cornish, N.J.; Liu, C. The construction and use of LISA sensitivity curves. \emph{Class. Quantum Grav.} \textbf{2019}, \emph{36}, 105011. \url{https://doi.org/10.1088/1361-6382/ab1101}.

\bibitem{LISA2017} Amaro-Seoane, P.;  Audley, H.; Babak, S.; Baker, J.; Barausse, E.; Bender, P.; Berti, E.; Binetruy, P.; Born, M.; Bortoluzzi, D.; %
 et al. Laser Interferometer Space Antenna.  \emph{arXiv} \textbf{2017}, arXiv:1702.00786. \url{https://doi.org/10.48550/arXiv.1702.00786}.

\bibitem{LindblomOwenBrown2008} Lindblom, L.; Owen, B.J.; Brown, D.A. Model waveform accuracy standards for gravitational wave data analysis. \emph{Phys. Rev. D} \textbf{2008}, \emph{78}, 124020. 
 \url{https://doi.org/10.1103/PhysRevD.78.124020}.

\bibitem{PrattenEtAl2021} Pratten, G.; Garc\'ia-Quir\'os, C.; Colleoni, M.; Ramos-Buades, A.; Estell\'es, H.; Mateu-Lucena, M.; Jaume, R.; Haney, M.; Keitel, D.; Thompson, J.E.; et al. Computationally efficient models for the dominant and subdominant harmonic modes of precessing binary black holes. \emph{Phys. Rev. D} \textbf{2021}, \emph{103}, 104056. \url{https://doi.org/10.1103/PhysRevD.103.104056}.

\bibitem{RamosBuadesEtAl2023} Ramos-Buades, A.; Buonanno, A.; Estell\'es, H.; Khalil, M.; Mihaylov, D.P.; Ossokine, S.; Pompili, L.; Shiferaw, M. SEOBNRv5PHM: Next generation of accurate and efficient multipolar precessing-spin effective-one-body waveforms for binary black holes. \emph{Phys. Rev. D} \textbf{2023}, \emph{108}, 124037. \url{https://doi.org/10.1103/PhysRevD.108.124037}.

\bibitem{TapaiEtAl2018} T\'apai, M.; Pint\'er, V.; Tarj\'anyi, T.; Keresztes, Z.; Gergely, L.\'A. Investigating the poor match among different precessing gravitational waveforms. \emph{Universe} \textbf{2018}, \emph{4}, 56. \url{https://doi.org/10.3390/universe4030056}.

\bibitem{LyuEtAl2024} Lyu, Z.; LaHaye, M.; Yang, H.; Bonga, B. Probing spin-induced quadrupole moments in precessing compact binaries. \emph{Phys. Rev. D} \textbf{2024}, \emph{109}, 064081. \url{https://doi.org/10.1103/PhysRevD.109.064081}.

\bibitem{LISAConsortium2025} LISA Consortium Waveform Working Group; Afshordi, N.; Akçay, S.; Amaro Seoane, P.; Antonelli, A.; Aurrekoetxea, J.C.; Barack, L.; Barausse, E.; Benkel, R.; Bernard, L.; %
 et al. Waveform modelling for the Laser Interferometer Space Antenna. \emph{Liv. Rev. Relativ.} \textbf{2025}, \emph{28}, 9. \url{https://doi.org/10.1007/s41114-025-00056-1}.

\bibitem{Morras2026} Morras, G.; Pratten, G.; Schmidt, P.; Buonanno, A. Post-Newtonian inspiral waveform model for eccentric precessing binaries with higher-order modes and matter effects. \emph{Phys. Rev. D} \textbf{2026}, \emph{114}, 044032. \url{https://doi.org/10.1103/lxtg-6psv}.

\bibitem{DamourNagarTrias2011} Damour, T.; Nagar, A.; Trias, M. Accuracy and effectualness of closed-form, frequency-domain waveforms for nonspinning black hole binaries. \emph{Phys. Rev. D} \textbf{2011}, \emph{83}, 024006. \url{https://doi.org/10.1103/PhysRevD.83.024006}.

\bibitem{VarmaEtAl2021} Varma, V.; Mould, M.; Gerosa, D.; Scheel, M.A.; Kidder, L.E.; Pfeiffer, H.P. Up-down instability of binary black holes in numerical relativity. \emph{Phys. Rev. D} \textbf{2021}, \emph{103}, 064003. \url{https://doi.org/10.1103/PhysRevD.103.064003}.

\bibitem{GerosaEtAl2023} Gerosa, D.; Fumagalli, G.; Mould, M.; Cavallotto, G.; Padilla Monroy, D.; Gangardt, D.; De Renzis, V. Efficient multi-timescale dynamics of precessing black-hole binaries. \emph{Phys. Rev. D} \textbf{2023}, \emph{108}, 024042. \url{https://doi.org/10.1103/PhysRevD.108.024042}.
\end{thebibliography}
\end{document}